\documentclass[12pt]{article}
\usepackage{amsthm}
\usepackage{amsmath}
\usepackage{ bbold }
\usepackage[colorlinks,citecolor=blue,urlcolor=blue,filecolor=blue,backref=page]{hyperref}
\usepackage{graphicx}
\usepackage[english]{babel}
\usepackage[utf8]{inputenc}
\usepackage{amsmath}
\usepackage{amssymb}
\usepackage{graphicx}
\usepackage{hyperref}
\usepackage{graphicx}
\usepackage{comment}
\usepackage{mathrsfs}
\usepackage{amsthm}
\usepackage{bbm}
\usepackage{scalerel}
\usepackage{rotating}
\usepackage{lscape}
\usepackage{subfigure}
\usepackage[colorinlistoftodos]{todonotes}
\usepackage{lipsum}
\usepackage{comment}
\usepackage[round]{natbib}
\usepackage[colorlinks,citecolor=blue,urlcolor=blue,filecolor=blue,backref=page]{hyperref}
\usepackage{graphicx}
\newtheorem{theorem}{Theorem}
\newtheorem{corollary}{Corollary}
\newtheorem{proposition}[theorem]{Proposition}
\newtheorem{lemma}[theorem]{Lemma}

\numberwithin{equation}{section}
\theoremstyle{plain}

\newcommand{\ignore}[1]{}

\usepackage{multirow}
\usepackage{amssymb}
\usepackage{graphicx}
\usepackage{comment}
\usepackage{mathrsfs}
\usepackage[colorinlistoftodos]{todonotes}
\usepackage{rotating}
\usepackage{lscape}
\usepackage{epstopdf}
\usepackage{natbib}
\usepackage{xcolor}

\usepackage{lipsum}
\def\undertilde#1{\mathord{\vtop{\ialign{##\crcr
				$\hfil\displaystyle{#1}\hfil$\crcr\noalign{\kern1.5pt\nointerlineskip}
				$\hfil\tilde{}\hfil$\crcr\noalign{\kern1.5pt}}}}}

\usepackage{subfigure}

\numberwithin{equation}{section}
\theoremstyle{plain}

\newcommand{\blind}{0}

\begin{document}
	\def\spacingset#1{\renewcommand{\baselinestretch}%
		{#1}\small\normalsize} \spacingset{1}

	\if0\blind
	{
		\title{Perturbed factor analysis: Accounting for group differences in exposure profiles}
		
		\author{Arkaprava Roy$^1$, Isaac Lavine$^2$, Amy Herring$^2$, David Dunson$^2$\\
			$^1$University of Florida, $^2$Duke University}
		\maketitle
	} \fi
	
	\if1\blind
	{
		\bigskip
		\bigskip
		\bigskip
		\begin{center}
			{\LARGE\bf }
		\end{center}
		\medskip
	} \fi
	
	\bigskip

	\begin{abstract}
		In this article, we investigate group differences in phthalate exposure profiles using NHANES data. Phthalates are a family of industrial chemicals used in plastics and as solvents.  There is increasing evidence of adverse health effects of exposure to phthalates on reproduction and neuro-development, and concern about racial disparities in exposure.  We would like to identify a single set of low-dimensional factors summarizing 
		exposure to different chemicals, while allowing differences across groups.  Improving on current multi-group additive factor models, we propose a class of Perturbed Factor Analysis (PFA) models that assume a common factor structure after perturbing the data via multiplication by a group-specific matrix.  Bayesian inference algorithms are defined using a matrix normal hierarchical model for the perturbation matrices.  The resulting model is just as flexible as current approaches in allowing arbitrarily large differences across groups but has substantial advantages that we illustrate in simulation studies. Applying PFA to NHANES data, we learn common factors summarizing exposures to phthalates, while showing clear differences across groups.
	\end{abstract}
	
	\noindent%

	\spacingset{1.45}

	\section{Introduction} 
	
	%Phthalate concentrations in blood or urine across different demographic groups are routinely collected by the National Health and Nutrition Examination Survey (NHANES). 
	
	Exposures to phthalates are ubiquitous. They are present in soft plastics, including vinyl floors, toys, and food packaging. Medical supplies such as blood bags and tubes contain phthalates. They are also found in fragrant products such as soap, shampoo, lotion, perfume, and scented cosmetics.  There is substantial interest in studying levels of exposure of people in different groups to phthalates, and in relating these exposures to health effects.
	This has motivated the collection of phthalate concentration data in urine in the National Health and Nutrition Examination Survey (NHANES) ({\url{https://wwwn.cdc.gov/nchs/nhanes}}). Excess levels of phthalates in blood/urine have been linked to a variety of health outcomes, including obesity \citep{zhang2014age, kim2014phthalate, benjamin2017phthalates} and birth outcomes \citep{bloom2019racial}. When they enter the body, phthalate parent compounds are broken down into different metabolites; assays measuring phthalate exposures target these different breakdown products. As these chemicals are often moderately to highly correlated, it is common to identify a small set of underlying factors
	(for example \cite{weissenburger2017principal}). Epidemiologists are interested in interpreting these factors and in using them in analyses relating exposures to health outcomes. 
	
	However, many research groups have noticed a tendency to estimate very different factors in applying factor analysis to similar datasets or groups of individuals. For example, \cite{maresca2016prenatal} examined data from three children's cohorts and noted marked differences in factor structure in one of them. \cite{james2017racial} found variation in phthalate exposure patterns of pregnant women by race. \cite{bloom2019racial} noted differences in both phthalate exposures and in associations with the birth outcomes by race. 
	Certainly, we would like to allow for possible differences in exposures across groups; indeed, studying such differences is one of our primary interests.  Such differences may relate to questions of environmental justice and may partly explain differences with ethnicity in certain health outcomes.  However, even though the levels and specific sources of phthalate exposures may vary across groups, there is no reason to suspect that the fundamental relationship between levels of metabolites and latent factors would differ.  Such differences in the factor structure are more likely to arise due to statistical uncertainty and sensitivity to slight differences in the data and can greatly complicate inferences on similarities and differences across groups.  
	
	%In order to make more generalizable health recommendations, it is important to understand typical population exposure patterns generalizable to broader populations in addition to understanding how those patterns may vary across groups of interest. %For reliable interpretability and generalizability, it is important to reduce this brittleness and may be desirable to have a method that can produce a single set of factors that hold across groups. %There should be some common factors underlying these chemical exposure levels in order for the factors to be interpretable. However, even a small amount of inter-group variation might make the estimation of those common factors problematic. Similar challenges arise whenever multiple batches of data are collected under the same experimental conditions. Even with small changes across the batches, the factors might end up being substantially different. 
	
	To set the stage for discussing factor modeling of multi-group data, consider a typical factor model for $p$-variate data $Y_i = (Y_{i1},\ldots,Y_{ip})^T$ from a single group:  
	\begin{eqnarray}
		Y_i = \Lambda \eta_i + \epsilon_i,\quad \eta_i \sim \textrm{N}(0,I_k),\quad \epsilon_i \sim \textrm{N}(0,\Sigma), \label{eq:FA} 
	\end{eqnarray}
	where it is assumed that data are centered prior to analysis, $\eta_i = (\eta_{i1},\ldots,\eta_{ik})^T$ are latent factors, $\Lambda$ is a $p \times k$ factor loadings matrix, and $\Sigma = \mbox{diag}(\sigma_{1}^2,\ldots,\sigma_{p}^2)$ is a diagonal matrix of residual variances. Under this model, marginalizing out the latent factors 
	$\eta_i$ induces the covariance $H = \mbox{cov}(Y_i)= \Lambda\Lambda^T + \Sigma$. 
	
	To extend \eqref{eq:FA} to data from multiple groups, most of the focus has been on decomposing the covariance into shared and group-specific components.  A key contribution was JIVE (Joint and Individual Variation Explained), which identifies a low-rank covariance structure that is shared among multiple data types collected for the same subject \citep{lock2013joint, feng2015non, feng2018angle}.  In a Bayesian framework, \cite{roy2019bayesian} proposed TACIFA (Time Aligned Common and Individual Factor Analysis) for a related problem. More directly relevant to our phthalate application are the multi-study factor analysis (MSFA) methods of \cite{de2018bayesian, de2019multi}.  These approaches replace $\Lambda \eta_i$ in (\ref{eq:FA}) with an additive expansion containing shared and group-specific components.  \cite{de2018bayesian} implement a Bayesian version of MSFA (BMSFA), while \cite{de2019multi} develop a frequentist implementation.  \cite{kim2018meta} propose a related approach using PCA.  
	
	The above methods are very useful in many applications but do not address our goal of improving inferences in our phthalate application by obtaining a common factor representation that holds across groups.  In addition, we find that additive expansions can face issues with weak identifiability - the model can fit the data well by decreasing the contribution of the shared component and increasing that of the group-specific components; this issue can lead to slower convergence and mixing rates for sampling algorithms for implementing BMSFA and potentially higher errors in estimating the component factor loadings matrices.

	We aim to identify a single set of phthalate exposure factors under the assumption that the data in different groups can be aligned to a common latent space via multiplication by a perturbation matrix. We represent the perturbed covariance in group $j$ as $Q_j\Sigma Q_j^{T}$, where $\Sigma$ is the common covariance, and $Q_j$ is the group-specific perturbation matrix. As in the common factor model in \eqref{eq:FA}, the overall covariance $\Sigma$ can be decomposed into a component characterizing shared structure and a 
	residual variance. The utility of the perturbation model also extends beyond multi-group settings. In the common factor model, the error terms only account for additive measurement error. We can obtain robust estimates of factor loadings from data in a single group by allowing for observation-specific perturbations. This accounts for both multiplicative and additive measurement error. In this case, we define separate $Q_i$'s for each data vector $Y_i$. Here $Q_i$'s are multiplicative random effects with mean $I_p$. Thus, $E(Q_iY_i)=E(Y_i)$ and the covariance structure on $Q_i$ would determine the variability of $Q_iY_i$.

	We take a Bayesian approach to inference using Markov chain Monte Carlo (MCMC) for posterior sampling, related to \cite{de2018bayesian} but using our Perturbed Factor Analysis (PFA) approach instead of their additive BMSFA model.  Model (\ref{eq:FA}) faces well known issues with non-identifiability of the loadings matrix $\Lambda$ \citep{seber2009multivariate,lopes2004bayesian,rovckova2016fast,fruehwirth2018sparse}; this non-identifiability problem is inherited by multiple group extensions such as BMSFA.  It is very common in the literature to run MCMC ignoring the identifiability problem and then post-process the samples. \cite{mcparland2014clustering, assmann2016bayesian} obtain a post-processed estimate by solving an orthogonal Procrustes problem, but without uncertainty quantification (UQ).  \cite{roy2019bayesian} post-process the entire MCMC chain iteratively to draw inference with UQ. \cite{lee2018mixtures} address non-identifiability by giving the latent factors non-symmetric distributions, such as half-$t$ or generalized inverse Gaussian. 
	We instead choose heteroscedastic latent factors in the loadings matrix and post-process based on \cite{roy2019bayesian}.  We find this approach can also improve accuracy in estimating the covariance, perhaps due to the flexible shrinkage prior that is induced.  
	
	The next section describes the data and the model in detail. In Section~\ref{prior}, prior specifications are discussed. Our computational scheme is outlined in Section~\ref{compute}. We study the performance of PFA in different simulation setups in Section~\ref{sim}.
	Section~\ref{real} applies PFA to NHANES data to infer a common set of factors summarizing phthalate exposures, while assessing differences in exposure profiles between ethic groups.  We discuss extensions of our NHANES analysis in Section~\ref{discuss}.
	
	\begin{table}[htbp]
		\centering
		\caption{NHANES data comparison across ethnic groups in terms of number of participants and average levels of different phthalate chemicals, with standard deviations in brackets.}
		\resizebox{.8\textwidth}{!}{
			\begin{tabular}{rrrrrr}
				\hline
				& Mex & OH & N-H White & N-H Black & Other/Multi \\ 
				\hline
				Number of participants & 566 & 293 & 1206 & 516 & 168 \\ 
				MnBP & 3.85 & 4.31 & 3.56 & 4.21 & 3.76 \\ 
				& (1.96) & (4.96) & (2.17) & (1.81) & (2.11) \\ 
				MiBP & 2.88 & 3.21 & 2.52 & 3.47 & 2.79 \\ 
				& (1.36) & (1.70) & (1.20) & (1.67) & (1.21) \\ 
				MEP & 8.32 & 9.51 & 6.91 & 11.16 & 7.33 \\ 
				& (6.64) & (6.99) & (5.43) & (9.38) & (7.57) \\ 
				MBeP & 2.79 & 2.78 & 2.70 & 3.06 & 2.56 \\ 
				& (1.58) & (1.61) & (1.67) & (1.77) & (1.68) \\ 
				MECPP & 4.80 & 4.55 & 4.16 & 4.36 & 4.34 \\ 
				& (3.68) & (2.68) & (2.40) & (2.38) & (2.64) \\ 
				MEHHP & 3.83 & 3.73 & 3.45 & 3.79 & 3.57 \\ 
				& (3.01) & (2.59) & (2.25) & (2.26) & (2.45) \\ 
				MEOHP & 3.11 & 3.00 & 2.79 & 3.05 & 2.86 \\ 
				& (2.41) & (1.92) & (1.69) & (1.71) & (1.86) \\ 
				MEHP & 1.58 & 1.59 & 1.33 & 1.61 & 1.58 \\ 
				& (1.23) & (1.29) & (0.89) & (0.99) & (1.31) \\ 
				\hline
		\end{tabular}}
		\label{demo}
	\end{table}

	\section{Data description and modeling}
	\label{model}
	
	NHANES is a population representative study collecting detailed individual-level data on chemical exposures, demographic factors and health outcomes.  Our interest is in using NHANES to study disparities across ethnic groups in exposures.
	Data are available on chemical levels in urine recorded from 2009 to 2013 for 2749 individuals. We consider eight phthalate metabolites, Mono-(2-ethyl)-hexyl (MEHP), Mono-(2-ethyl-5-oxohexyl) (MEOHP), Mono-(2-ethyl-5-hydroxyhexyl) (MEHHP), Mono-2-ethyl-5-carboxypentyl (MECPP), Mono-benzyl (MBeP), Mono-ethyl (MEP),Mono-isobutyl (MiBP), and Mono-n-butyl (MnBP), and measured in participants identifying with racial and ethnic groups: Mexican American (Mex), Other Hispanic (OH), Non-Hispanic White (N-H White), Non-Hispanic Black (N-H Black) and Other Race (Other) which includes multi-racial. Previous work has shown differences across ethnic groups in patterns of use of products that contain phthalates \citep{taylor2018associations} and in measured phthalate concentrations \citep{james2017racial}. Recent work has also indicated exposure effects themselves may vary across ethnic groups \citep{bloom2019racial}, but this may be difficult to disentangle if summaries of exposure lack sufficient robustness and generalizability.

\begin{figure}[htbp]
		\centering
		\includegraphics[width = 1\textwidth]{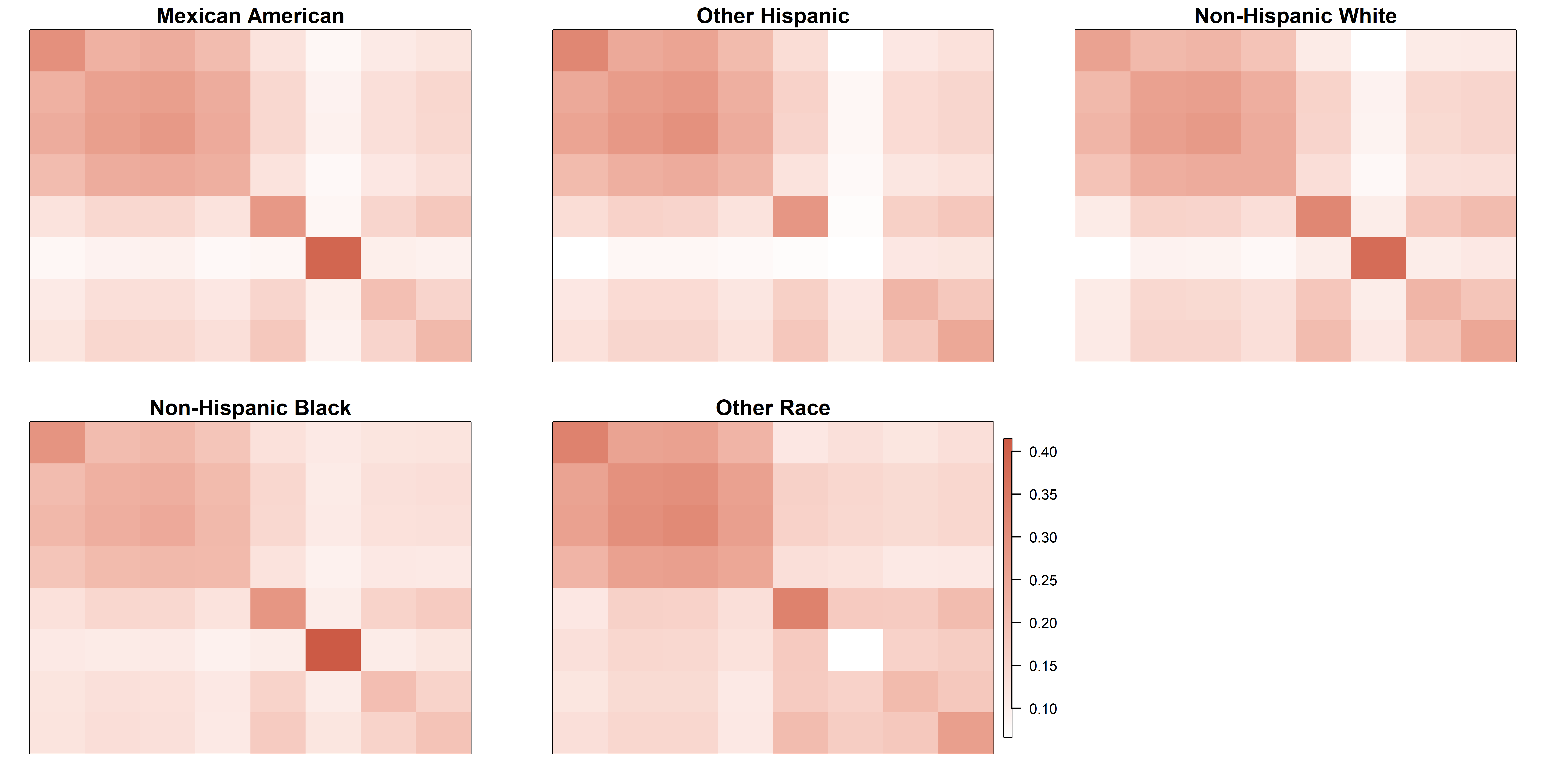}
		\caption{Covariance matrices of the phthalate chemicals across different ethnic groups for NHANES data.}
		% DD - it is implied that a common color gradient holds across the panels
		\label{realcov}
	\end{figure}

	We summarize the average level of each chemical in each group in 
	Table~\ref{demo}. In Table~\ref{htsqr}, we compute the Hotelling $T^2$ statistic between each pair of groups. As noted in \cite{bloom2019racial}, exposure levels were generally higher among non-whites. For each phthalate variable, we also fit a one-way ANOVA model to assess differences across the groups taking Mexican-Americans as the baseline. The results are included in Tables 2-6 in the supplementary materials. For the majority of phthalates, concentrations are lowest among non-Hispanic whites. We plot group-specific covariances in Figure~\ref{realcov}. It is evident that there is shared structure with some differences across the groups.	
	
	\begin{table}[htbp]
		\centering
		\caption{Hotelling $T^2$ statistic between phthalate levels for each pair of groups in NHANES.}
		\begin{tabular}{rrrrrr}
			\hline
			& Mex & OH & N-H White & N-H Black & Other/Multi \\ 
			\hline
			Mex & 0.00 & 31.84 & 108.59 & 191.93 & 28.04 \\ 
			OH & 31.84 & 0.00 & 156.45 & 63.48 & 28.69 \\ 
			N-H White & 108.59 & 156.45 & 0.00 & 409.64 & 55.51 \\ 
			N-H Black & 191.93 & 63.48 & 409.64 & 0.00 & 101.90 \\ 
			Other/Multi & 28.04 & 28.69 & 55.51 & 101.90 & 0.00 \\ 
			\hline
		\end{tabular}
		\label{htsqr}
	\end{table}
	
	\begin{figure}[htbp]
		\centering
		\includegraphics[width = 1\textwidth]{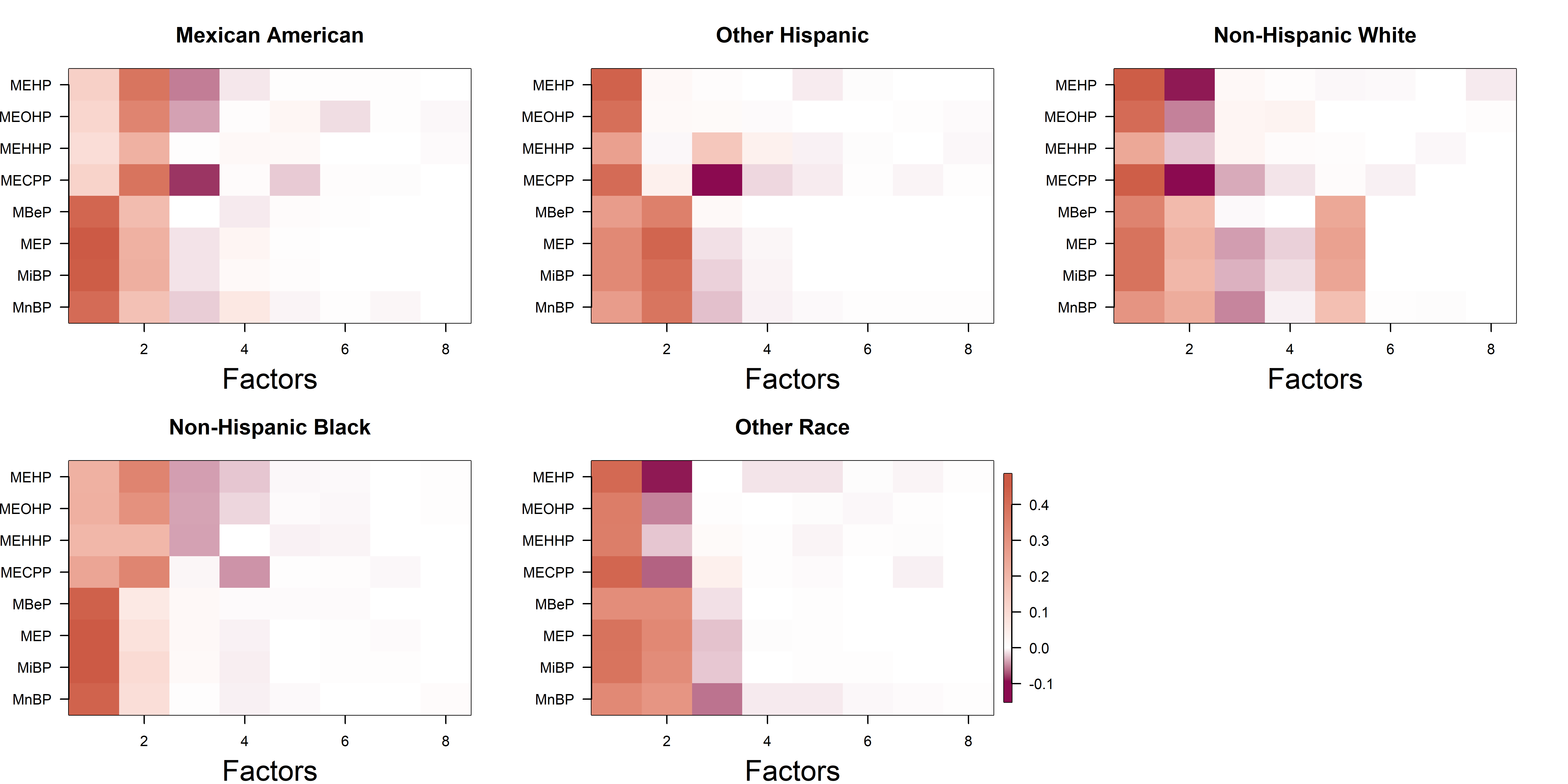}
		\caption{Estimated loadings matrices for different ethnic groups in NHANES based on applying separate Bayesian factor analyses using the approach of \cite{bhattacharya2011sparse}.}
		\label{realCF}
	\end{figure}

	Fitting the common factor model~\eqref{eq:FA} separately for each group, we obtain noticeably different loadings structures as depicted in Figure~\ref{realCF}. Throughout the article, including for Figures \ref{realcov} and \ref{realCF}, we plot matrices using the {\tt mat2cols()} function for R package {\tt IMIFA} \citep{murphy2018infinite,IMIFApk}.  This function standardizes the color scale across the different matrices.  We choose color palettes coral3 and deeppink4 from R package {\tt RColorBrewer} \citep{Rcolorbrewer} for positive and negative entries, respectively. Zero entries are white, and the color gradually changes as the value moves away from zero. In the next subsection, we propose a novel approach for
	multi-group factor analysis to identify common phthalate factors.

	%\section{Model}
	
	\iffalse

	A standard latent factor model with Gaussian errors models the covariance of independent observations $Y_i$, $i \in 1:n$ as:
	\begin{equation}
		Y_i = \Lambda \eta_i + \epsilon \qquad \epsilon \sim \textrm{N}(0, \Sigma_1),
	\end{equation}
	where  $\Lambda$ is the $p$ by $k$ dimensional factor loadings matrix, and $\Sigma_1$ is a diagonal matrix with residual noise. The latent factors for person $i$ are $\eta_i$, distributed $\eta_i \sim \textrm{N}(0, I_k)$. Marginally, this implies that $Y_i \sim \textrm{N}(0, \Sigma)$, where $\Sigma = \Lambda \Lambda^T + \Sigma_1$. %The purpose of a factor model is to discover the $k$ factor loadings in $\Lambda$ which can explain the covariance structure in $\Sigma$.
	\fi
	
	\subsection{Multi-group model}
	We have data from multiple groups, with the groups corresponding to individuals in different racial and ethnic categories in NHANES.  Each $p$-dimensional  mean-centered response $Y_{ij}$, belonging to group $G_j$, for $j \in 1:J$ and $i \in 1:n_j$, is modeled as:
	\begin{align}
		Q_jY_{ij}=&\Lambda\eta_{ij}+\epsilon_{ij},\nonumber\\
		Q_j\sim & \mathrm{MN}_{p\times p}(I_p, U, V),\quad \eta_{ij}\sim \textrm{N}(0, E),\nonumber\\
		\epsilon_{ij}\sim & \textrm{N}(0, \Sigma).\label{GMSEM}
	\end{align}
	The perturbation matrices $Q_j$ are of dimension $p\times p$, and follow a matrix normal distribution, with isotropic covariance $U = V = \alpha I_p$.  The latent factors $\eta_{ij}$ are heteroscedastic, so that $E$ is diagonal with non-identical entries such that $E=\textrm{diag}(e_{1},\ldots,e_{k})$ with $k$ factors in the model and $\Sigma=\textrm{diag}(\sigma_{1},\ldots,\sigma_{p})$. We discuss advantages of choosing heteroscedastic latent factors in more detail in Section~\ref{splmodel}. After integrating out the latent factors, observations are marginally distributed as $Y_{ij} \sim \textrm{N}(0, Q_j^{-1}[\Lambda E\Lambda^T+\Sigma](Q_j^{-1})^T)$.  If we write $Q_j^{-1} = I_p + \Psi_j$, then $\Lambda$ is the shared loadings matrix, and $\Psi_j\Lambda$ is a group-specific loadings matrix. 
	%We assume that $U$ and $V$ are diagonal matrices with equal diagonal entries to impose similar level of pertubation for all the components. 
	We can quantify the magnitude of perturbation as $\|\Psi_j\|_F$, where $\|\cdot\|_F$ stands for the Frobenius norm. For identifiability of $Q_j$'s, we consider $Q_1=I_p$ and $n_1\geq 2$. We call our model Perturbed Factor Analysis (PFA).

	\subsubsection{Model properties}
	\label{modelprop}
	
	Let $\Omega_1$ and $\Omega_2$ be two positive definite (p.d.) matrices. Then there exist non-singular matrices $A$ and $B$, where $\Omega_1=AA^T$ and $\Omega_2=BB^T$. By choosing $E=AB^{-1}$, we have $\Omega_1=E\Omega_2E^T$. If the two matrices $\Omega_1$ and $\Omega_2$ are close, then $E$ will be close to the identity. However, $E$ is not required to be symmetric. In our multi-group model in \eqref{GMSEM}, the $Q_j$s allow for small perturbations around the shared covariance matrix $H=\Lambda E\Lambda^T+\Sigma$. We define the following class for the $Q_j$'s,
	
	$$
	\mathcal{C}_{\epsilon}=\{Q:\|Q-I_p\|_F\leq \epsilon\}.
	$$
	
	The index $\epsilon$ controls the amount of deviation across groups. In \eqref{GMSEM} we define $U=V=\alpha I_p$, for some small $\alpha$, which we call the perturbation parameter. By choosing $U = V$ to be isotropic covariances, we impose uniform perturbation across all the rows and columns around $I_p$. However, the perturbation matrices themselves are not required to be symmetric.
	
	\begin{lemma}
		$P\big(Q_i\notin \mathcal{C}_{\epsilon}\big)\leq \exp\big(-\epsilon^2/2\alpha^2\big)$.
	\end{lemma}
	The proof follows from Chebychev's inequality. This result allows us to summarize the level of induced perturbation for any given $\alpha$. Using this lemma, we can show the following esult
	\begin{comment}
	\begin{corollary}
	If the true perturbation matrix $Q_{i0}$ is distributed MN$(I_p, \alpha I_p, \alpha I_p)$, then
	$$
	P(\|Q_{i0}-Q_{i}\|_{F}<2\epsilon)\geq 2\big(1-\exp\Large(-\epsilon^2/2\alpha^2\Large)\big).
	$$
	\end{corollary}
	where $Q_i$ is the estimated perturbation matrix. Thus, as the true $\alpha$ decreases, we become more confident in our estimate of the perturbation.
	\end{comment}
	
	%Above conditional distribution can be used to test significance of the difference between $Q_j$ and $Q_l$.
	\begin{align*}
		&\textrm{KL}(\textrm{N}(0, Q_j^{-1}[\Lambda E\Lambda^T+\Sigma](Q_j^{-1})^T), \textrm{N}(0, Q_l^{-1}[\Lambda E\Lambda^T+\Sigma](Q_l^{-1})^T))\\&\quad\lesssim {\big|\|Q_j^{-1}\|_F^2-\|Q_l^{-1}\|_F^2\big|}.
	\end{align*}
	
	Therefore the Kullback-Leibler divergence between the marginal distributions of any two groups $j$ and $l$ can be bounded by $\big|\|Q_j^{-1}\|_F^2-\|Q_l^{-1}\|_F^2\big|$ up to some constant. We define a divergence statistic between groups $j$ and $l$ as $d_{jl}=\sqrt{\frac{\big|\|Q_j^{-1}\|_F^2-\|Q_l^{-1}\|_F^2\big|}{p^2}}$. This generates a divergence matrix $D=(\{d_{jl}\})$, where larger $d_{jl}$ implies a greater difference between the two groups. 
	
	As in other factor modeling settings, it is important to carefully choose the number of factors.  This is simpler for PFA than for other multi-group factor analysis models, because we only need to select a single number of factors instead of separate values for shared and individual-specific components.  Indeed, we can directly apply methods developed in the single group setting
	\citep{bhattacharya2011sparse}. 
	Due to heteroscedastic latent factors, we can directly use the posterior samples of the loading matrices without applying factor rotation. Computationally, PFA tends to be faster than multi-group factor models such as BMSFA due to a smaller number of parameters, and better identifiability, leading to better mixing of Markov chain Monte Carlo (MCMC). 
	
	\subsubsection{Choosing the parameter \texorpdfstring{$\alpha$}{e} for multi-group data}
	\label{tune}
	
	The parameter $\alpha$ controls the level of perturbation across groups. 
	%In Section~\ref{sim}, we show that properly choosing $\alpha$ is necessary to obtain accurate estimates of the loading matrix $\Lambda$. 
	We use a cross-validation technique to choose $\alpha$ based on 10 randomly chosen 50-50 splits. We randomly divide each group $50/50$ into training and test sets ten times. Then for a range of $\alpha$ values, we fit the model on the training data and calculate the average predictive log-likelihood of the test set for all the ten random splits. After integrating out the latent factors, the predictive distribution is $Q_jY_{ij} \sim \textrm{N}(0, \Lambda E\Lambda^T+\Sigma)$. If there are multiple values of $\alpha$ with similar predictive log-likelihoods, then the smallest $\alpha$ is chosen. 
	
	Alternatively, we can take a fully Bayesian approach and put a prior on $\alpha$. We call this method Fully Bayesian Perturbed Factor Analysis (FBPFA). We see that FBPFA performs similarly to PFA in practice, but involves a slightly more complex MCMC implementation. PFA avoids sensitivity to the prior for $\alpha$ but requires the computational overhead of cross validation,
	ignores uncertainty in estimating $\alpha$, and 
	and can potentially be less efficient in requiring a holdout sample.

	\subsection{Measurement-error model}
	
	We can modify the multi-group model to obtain improved factor estimates in single group analyses by considering observation-level perturbations. Here we observe $Y_{ij}$'s for $j=1,\ldots,m_i$, which are $m_i$ many proxies of `true' observation $W_i$ with multiplicative measurement errors $Q_{ij}^{-1}$'s such that $W_{i}=Q_{ij}Y_{ij}$. The modified model is
	\begin{align}
		Q_{ij}Y_{ij}=&\Lambda\eta_{i}+\epsilon_{ij},\quad\epsilon_{ij}\sim \textrm{N}(0, \Sigma),\nonumber\\
		Q_{ij}\sim & \mathrm{MN}_{p\times p}(I_p, U, V)\quad \eta_{i}\sim \textrm{N}(0, E).\label{MSEM}
	\end{align}
	In this case, the $Q_{ij}$'s apply a multiplicative perturbation to each data vector. We have $Y_{ij}=Q_{ij}^{-1}W_i$ where $Q_{ij}^{-1}$ is a matrix. Thus, here the measurement errors are $U_{ij}=(Q_{ij}^{-1}-I_p)W_i$ and $E(U_{ij}|W_i)=0$. This model is different from the multiplicative measurement error model of \cite{sarkar2018bayesian}. In their paper, observations $Y_{ij}$'s are modeled as $Y_{ij}=W_i\circ U_{ij}$,  where $\circ$ denotes the element wise dot product and $U_{ij}$'s are independent of $W_i$ with $E(U_{ij})=1$. Thus, the measurement error in the $l$-$th$ component (i.e. $U_{ijl}$) is dependent on $W_{i}$ primarily through $W_{il}$. However, in our construction, the measurement errors are a linear function of the entire $W_i$. 
	
	This is a much more general setup than \cite{sarkar2018bayesian}. With this generality comes issues in identifying parameters in the distributions of $Q_{ij}$ and $Y_{ij}$. For simplicity, we again assume $U=V=\alpha I_p$.  In this case we have,
	\begin{align*}
		E(Q_{ij}Y_{ij})&=0\\
		V(Q_{ij}Y_{ij})&=E(V(Q_{ij}Y_{ij}|Q_{ij}))+V(E(Q_{ij}Y_{ij}|Q_{ij}))\\
		&=E(V(Q_{ij}Y_{ij}|Q_{ij}))+0=\alpha^2sI_p+H, \quad s=\sum_{j=1}^pH_{jj}
	\end{align*}
	Thus, only the diagonal elements of $H$ are not identifiable and the perturbation parameter $\alpha$ does not influence the dependence structure among the variables. Hence, with our heteroscedastic latent factors, we can still recover the loading structure. To tune $\alpha$, we can use the marginal distributions $Q_{ij}Y_{ij}\sim$N$(0, \alpha^2sI_p+H)$ to develop a cross validation technique when $m_i>1$ for all $i=1,\ldots, n$ as in Section~\ref{tune}. We split the data into training and testing sets. Then fit the model in the training set and
	find the $\alpha$ minimizing the predictive log-likelihood in the test set.  We can alternatively estimate $\alpha$ as in FBPFA by using a weakly informative prior concentrated on small values, while assessing sensitivity to hyperparameter choice.

	%Then some prior knowledge of the data generating mechanism would be helpful to determine the level of perturbation for setting the perturbation parameter $\alpha$. We can also consider the cross-validation technique to choose $\alpha$ or put a weakly informative inverse-gamma prior.
	
	\subsection{The special case \texorpdfstring{$Q_j=I_p$}{e} for all \texorpdfstring{$j$}{e}}
	\label{splmodel}
	For data within a single group without measurement errors (in the sense considered in the previous subsection),  we can modify PFA by taking $Q_j=I_p$ for all $j$. Then the model reduces to a traditional factor model with heteroscedastic latent factors:
	\begin{align}
		Y_i=&\Lambda\eta_{i}+\epsilon_{i},\quad \epsilon_{i}\sim \textrm{N}(0, \Sigma)\nonumber\\
		\eta_{i}\sim & \textrm{N}(0, E),\label{MFA}
	\end{align}
	where $E$ is assumed to be diagonal with non-identical entries. Integrating out the latent factors, the marginal likelihood is $Y_i \sim \textrm{N}(0, \Lambda E\Lambda^T+\Sigma)$. Except for the diagonal matrix $E$, the marginal likelihood is similar to the marginal likelihood for a traditional factor model. As $E$ has non-identical diagonal entries, the likelihood is no longer invariant under arbitrary rotations. For the factor model in (\ref{eq:FA}), $(\Lambda,\eta)$ and $(\Lambda R, R^T\eta)$ have equivalent likelihoods for any orthonormal matrix $R$. This is not the case in our model unless $R$ is a permutation matrix. Thus, this simple modification over the traditional factor model helps to efficiently recover the loading structure. This is demonstrated in Case 1 of Section~\ref{sim}. We also show that the posterior is weakly consistent in the supplementary materials.
	
	\section{Prior Specification}
	\label{prior}
	As in \cite{bhattacharya2011sparse}, we put the following prior on $\Lambda$ to allow for automatic selection of rank and easy posterior computation:
	
	$$\lambda_{lk}|\phi_{lk},\tau_{k}\sim \mathrm{N}(0,\phi_{lk}^{-1}\tau_{k}^{-1}),$$
	$$\phi_{lk}\sim \mathrm{Gamma}(\nu_1,\nu_1),\quad \tau_{k}=\prod_{i=1}^k\delta_{i}$$
	$$\delta_{1}\sim \mathrm{Gamma}(\kappa_{1}, 1),\quad\delta_{i}\sim \mathrm{Gamma}(\kappa_{2}, 1)\textrm{ for }i\geq 2.$$
	The parameters $\phi_{lk}$ control local shrinkage of the elements in $\Lambda$, whereas $\tau_k$ controls column shrinkage of the $k$-$th$ column.  We follow the guidelines of \cite{durante2017note} to choose hyperparameters that ensure greater shrinkage for higher indexed columns. 
	In particular, we let $\kappa_1=2.1$ and $\kappa_2=3.1$, which works well in all of our simulation experiments. 
	%{\color{red}Instead of the multiplicative gamma process, one can also consider the new class of cumulative shrinkage process prior, inspired from Sethuraman's stick-breaking construction of Dirichlet process \citep{legramanti2020bayesian}.}
	
	Suppose we initially choose the number of factors to be too large. The above prior will then tend to induce posteriors for $\tau_k^{-1}$ in the later columns that are concentrated near zero, leading to $\lambda_{lk} \approx 0$ in those columns.  The corresponding factors are then effectively deleted.  
	%In calculating summary statistics of the induced covariance, rank selection is not needed.  
	In practice, one can either leave the extra factors in the model, as they will have essentially no impact, or conduct a factor selection procedure by removing factors having all their loadings within $\pm \zeta$ of zero. Both approaches tend to have similar performance in terms of posterior summaries of interest.
	We follow the second strategy, motivated by our goal of obtaining a small number of interpretable factors summarizing phthalate exposures. In particular, we apply the adaptive MCMC procedure of \cite{bhattacharya2011sparse} with $\zeta = 1 \times 10^{-3}$.
	
	For the heteroscedastic latent factors, each diagonal element of $E$ has an independent prior:
	$$
	e_{i}\sim\text{IG}(u, 0.1)
	$$
	for some constant $u$. In our simulations, we see that $u$ has minimal influence on the predictive performance of PFA.  However, as $u$ increases, more shrinkage is placed on the latent factors.  We choose $u=10$ for most of our simulations. For the residual error variance $\Sigma$, we place a weakly informative prior on the diagonal elements:
	$$\sigma_{i}\sim\text{IG}(0.1,0.1).$$ In our simulations, a weakly informative {IG}(0.1,0.1) prior on $\alpha$ works well in terms of both predictive performance and estimation of the loading structure, including in single group analyses. 
	
	\section{Computation}
	\label{compute}
	Posterior inference is straightforward with a Gibbs sampler, because all of the parameters have conjugate full conditional distributions. For the model in~\eqref{GMSEM}, the full conditional of the perturbation matrix $Q_j$ is:
	$$\textrm{vec}(Q_j)|Y\sim \textrm{N}\big(\Gamma_j(V\otimes U)^{-1}\textrm{vec}(I_p), \quad \Gamma_j \big),$$
	where $\Gamma_j=\big[(V\otimes U)^{-1}+S_j\bigotimes H^{-1}\big]^{-1}$ and $S_j=\sum_{i}Y_{ij}Y_{ij}^T$. The notation $\bigotimes$ stands for Kronecker's product. The full conditionals for all other parameters are described in \cite{bhattacharya2011sparse}, replacing $Y_{ij}$ by $Q_jY_{ij}$. For the model in~\eqref{MSEM}, the full conditional of $Q_{ij}$ is: 
	$$\textrm{vec}(Q_{ij})|Y\sim \textrm{N}\big(\Gamma_{ij}(V\otimes U)^{-1}\textrm{vec}(I_p), \quad \Gamma_{ij} \big),$$
	where $\Gamma_{ij}=\big[(V\otimes U)^{-1}+S_{ij}\bigotimes H^{-1}\big]^{-1}$ and $S_j=Y_{ij}Y_{ij}^T$. Other parameters can again be updated using the results in \cite{bhattacharya2011sparse} replacing $Y_{ij}$ by $Q_{ij}Y_{ij}$. To sample the entire $\textrm{vec}(Q_{ij})$ or $\textrm{vec}(Q_{j})$ together, we need to invert a $p^2\times p^2$ matrix at each step. Instead we iteratively update the columns of $Q_{ij}$ or $Q_{j}$, which does not require matrix inversion when $U=V=\alpha I_p$. For simplicity in notation, we only show the update for the columns in $Q_j$ of the model in~\eqref{GMSEM}. The full conditional of the $l$-$th$ column in $Q_{j}$ is 
	$$
	Q_{j,l}|Y\sim \textrm{N}\big(M_lV_l, \ V_l \big),
	$$
	where $V=1/(\sum_{k\in G_j}Y_{l,k}^2/\textrm{diag}(\Sigma) + 1/\alpha)$ and 
	$M_l=-\sum_{k\in G_j}(Q_{j,-l}Y_{-l,k}-\Lambda\eta_{k})Y_{l,k}/\Sigma+g_l/\alpha$, 
	%and $V=1/(\sum_{k\in G_j}Y_{l,k}^2/\textrm{diag}(\Sigma) + 1/\alpha)$. 
	with $\Sigma$ the diagonal error covariance matrix. Here $Q_{j,-l}$ denotes the $p\times (p-1)$ dimensional matrix removing the $l$-$th$ column from $Q_{j}$. Similarly $Y_{-l,k}$ denotes the response of the $k$-$th$ individual from group $G_j$, removing the $l$-$th$ variable, and $g_l$ the $p$-dimensional vector with one in the $l$-th entry. If we put a prior on $\alpha$, the posterior distribution of $\alpha$ is given by IG$(0.1+(J-1)p^2, 0.1+\sum_j\|Q_{j}-I_p)^2\|_F^2)$ for the model in~\eqref{GMSEM} and for the model in~\eqref{MSEM}, the posterior distribution of $\alpha$ is IG$(0.1+p^2\sum_im_i, 0.1+\sum_{ij}\|Q_{ij}-I_p)^2\|_F^2)$. The posterior distribution of $\sigma_{1l}^{2}$ is IG$(0.1+n/2 ,0.1+\sum_{j=1}^J\sum_{k\in G_j}(Q_{j}(l,)Y_{kj}-\lambda_{l,}\eta_{kj})^2/2)$, where $Q_{j}(l,)$ denotes the $l$-$th$ row of $Q_j$. The posterior distribution of $\sigma_{2l}^{2}$ is IG$(0.1+n/2 ,0.1+\sum_{j=1}^J\sum_{k\in G_j}\eta_{l,kj}^2/2)$.

	\iffalse
	\begin{proof}
		We know from the properties of matrix normal that. Then the conditional distribution of $Q_j$ given the data are 
		
		$$
		\textrm{vec}(Q_j)|Y\sim \textrm{N}\big(\Gamma_j(V\otimes U)^{-1}\textrm{vec}(I_p), \quad \Gamma_j \big),
		$$
		where $\Gamma_j=\big[(V\otimes U)^{-1}+S_j\bigotimes\Sigma^{-1}\big]^{-1}$ with $S_j=\sum_{i}Y_{ij}Y_{ij}^T$.
	\end{proof}
	\fi
	
	%Using this distribution, we can use KL divergence to test the difference between $Q_j$ and $Q_l$.
	
	\section{Simulation Study}
	\label{sim}
	
\begin{figure}[htbp]
		\centering
		\includegraphics[width = 0.8\textwidth]{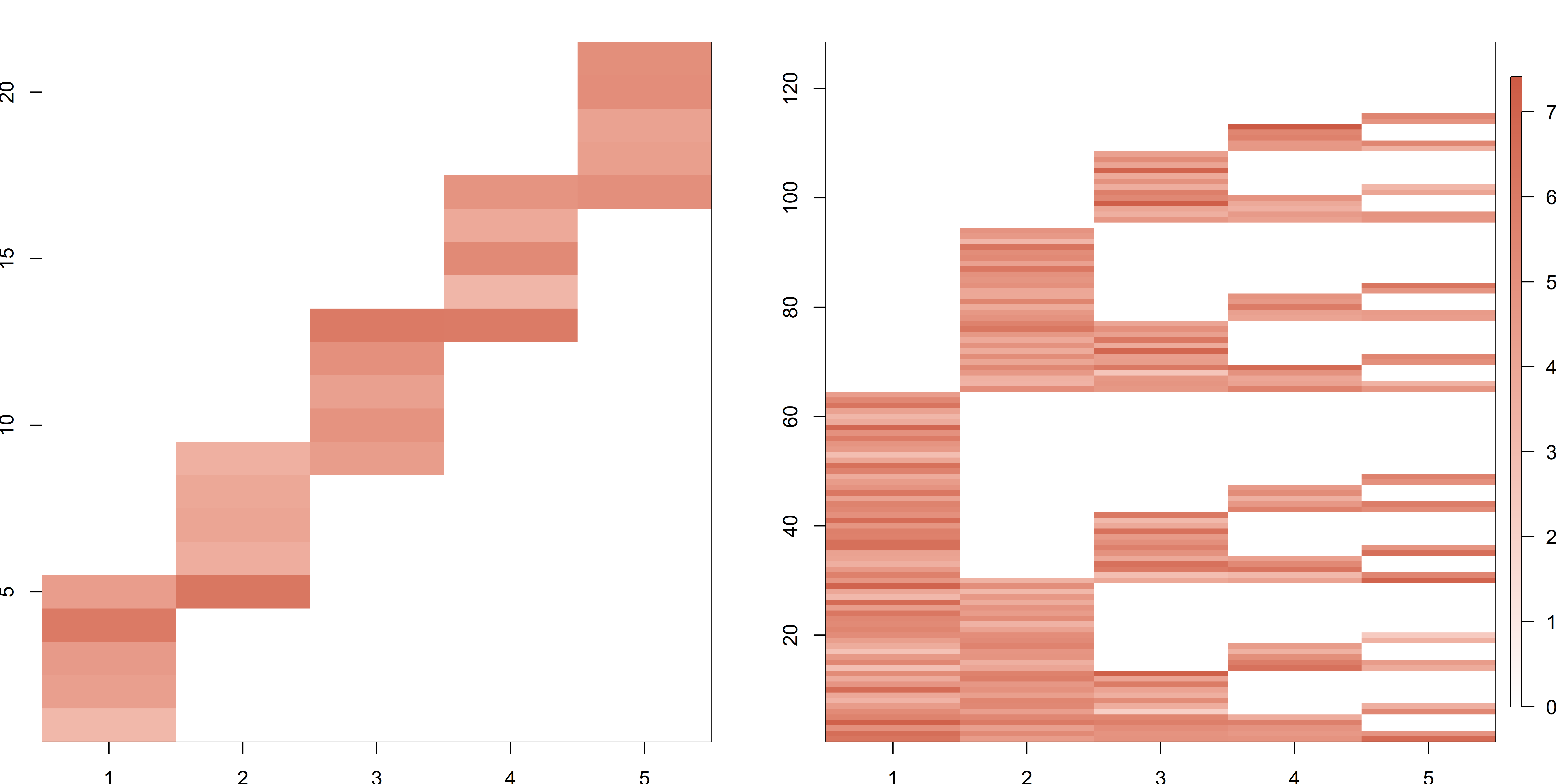}
		\caption{Simulation study true loading matrices of dimension $2 5$ (Loading 1) and $128\times 5$ (Loading 2), respectively.}
		\label{true}
	\end{figure}

	In this section, we study the performance of PFA in various simulation settings. As ground truth, we consider the two loading matrices given in Figure~\ref{true}. These are similar to ones considered in \cite{rovckova2016fast}. R code to generate the data is provided in the supplementary materials. The two matrices both have 5 columns, 
	the first has 21 rows, and the second has 128 rows. We compare the estimated loading matrices for different choices of the perturbation parameter  $\alpha$ and the shape parameter $u$, controlling the level of shrinkage on the diagonal entries of $E$.
	%We use the {\tt mat2cols()} function from \cite{murphy2018infinite, IMIFApk} to plot the loading matrices for better visualization.

	For the single group case, we compare with the method of \cite{bhattacharya2011sparse} (B\&D), which corresponds to the special case of our approach that fixes the perturbation matrices and latent factor covariances equal to the identity. A point estimate of the loading matrix for B\&D is calculated by post-processing the posterior samples. We use the algorithm of \cite{assmann2016bayesian} to rotationally align the samples of $\Lambda$, as in \cite{de2018bayesian}. In contrast, our method uses the post-processing algorithm of \cite{roy2019bayesian}. For the multi-group case, we compare our estimates with Bayesian Multi-Study Factor Analysis (BMSFA). We use the {\tt MSFA} package at {\url{https://github.com/rdevito/MSFA}}. 
	%Due to the non-identifiability of the perturbation parameter $\alpha$, we do not report its posterior mean for FBPFA.
	
	We compare the different methods using predictive log-likelihood. Simulated data are randomly divided $50/50$ into training and test sets. After fitting each model on the training data, we calculate the predictive log-likelihood of the test set. All methods are run for $7000$ iterations of the Gibbs sampler, with $2000$ burn-in samples. Each simulation is replicated $30$ times.
	
	\begin{figure}[htbp]
		\centering
		\subfigure[]{\label{fig:a}\includegraphics[width=125mm]{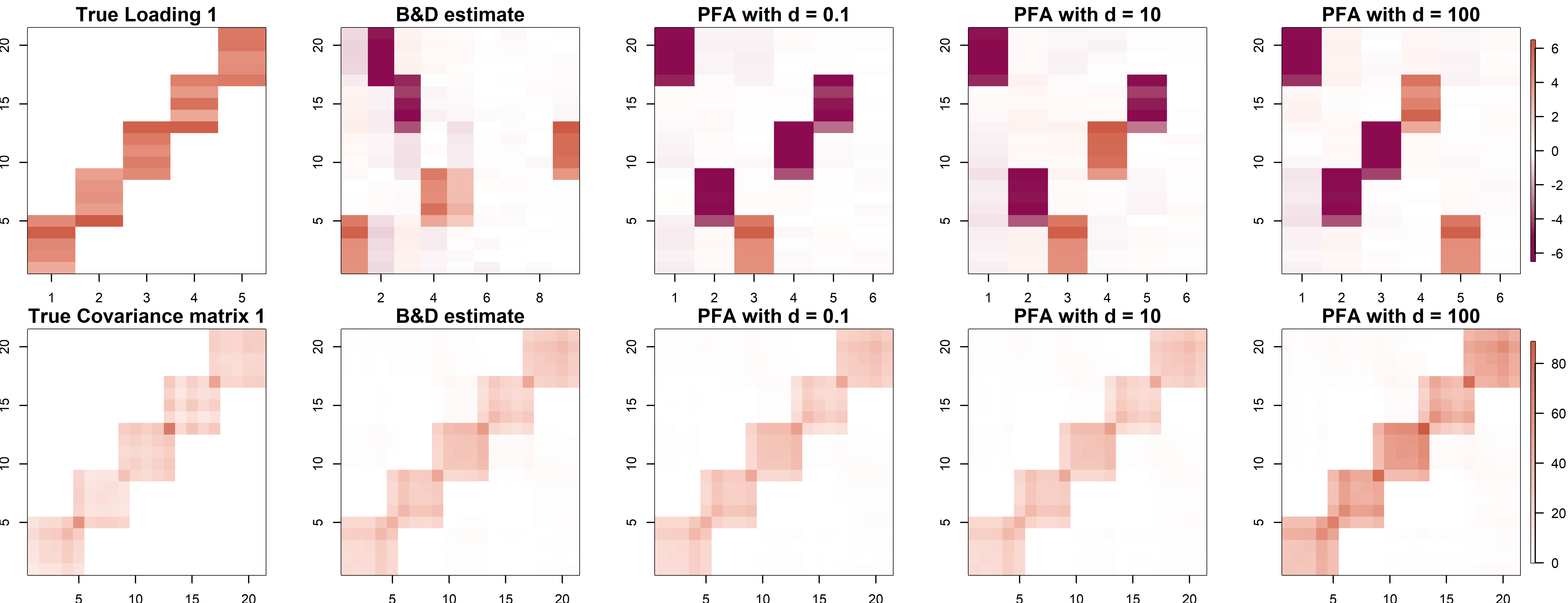}}
		\subfigure[]{\label{fig:b}\includegraphics[width=125mm]{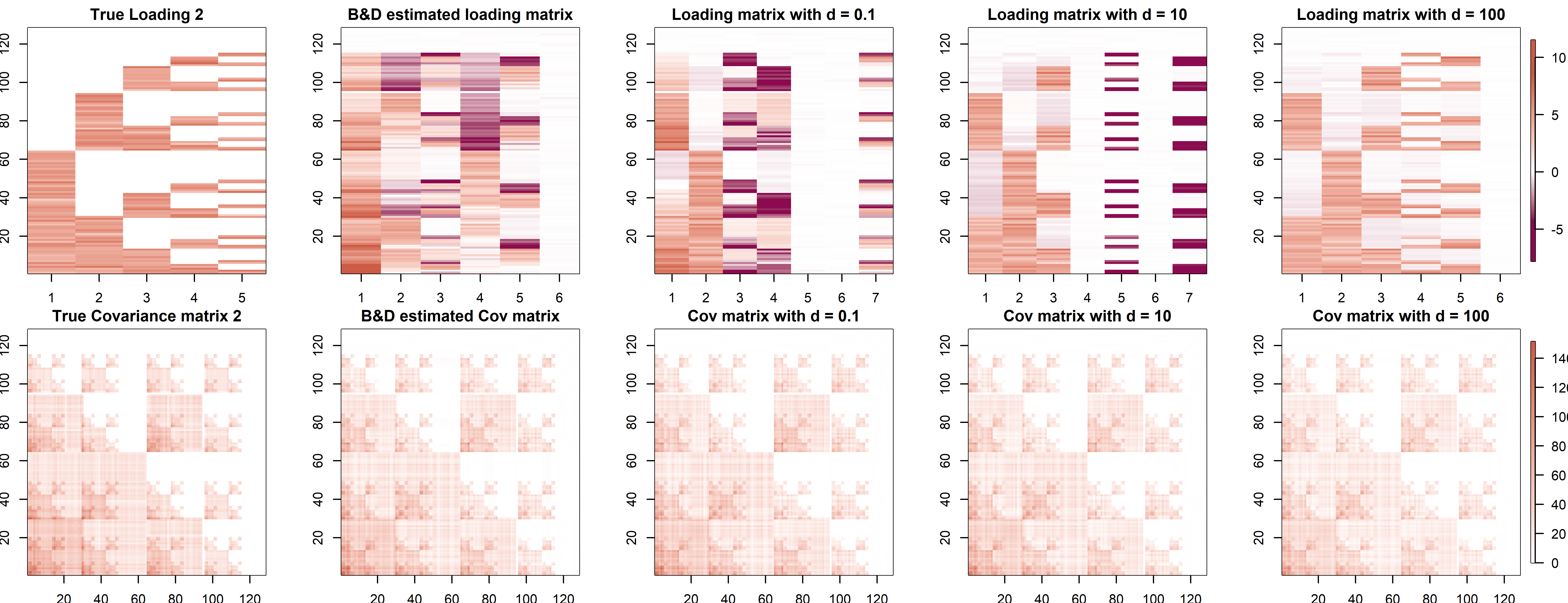}}
		\caption{Comparison of true and estimated loading matrices and covariance matrices for case $1$. Comparison of B\&D FA with our method for different choices of $u$.  Results for loading matrix 1 are in (a) and 2 in (b). For all the cases, we fix $Q_j=I$.The color gradient added in the last image of each row holds for all the images in that row.}
		\label{nopert}
	\end{figure}

	\subsection{Case 1: Single group, \texorpdfstring{$Q_j = I_p$}{e}}
	\label{simG1}
	
	We first consider single group factor analysis. Starting with the two loading matrices in Figure~\ref{true}, we simulate latent factors from $\textrm{N}(0, I_5)$, and generate datasets of $500$ observations, with residual variance $\Sigma = I_5$. We compare B\&D with our method when $Q_j=I_5$, so the only adjustment is to use heteroscedastic latent factors. Note that the simulated latent factors actually have identical variances.
	
	\begin{table}[htbp]
		\centering
		\caption{MSE of estimated covariance matrix for loading matrix $1$, across different values of $u$}
		\begin{tabular}{rr}
			\hline
			$u$ & MSE \\ 
			\hline
			0.1 & 1.04 \\
			10 & 1.32\\
			100 & 1.38\\
			%1000& 1.37 &    \\
			\hline
		\end{tabular}
		\label{cov1}
	\end{table}
	
	In Tables~\ref{cov1} and~\ref{cov2}, we compare methods by MSE of the estimated versus true covariance matrix. For the loading matrix $1$, B\&D had an MSE of $2.59$, which is dominated by our method across a range of values of $u$.
	
	\begin{table}[htbp]
		\centering
		\caption{MSE of estimated covariance matrix for loading matrix $2$, across different values of $u$}
		\begin{tabular}{rr}
			\hline
			$u$ & MSE  \\ 
			\hline
			0.1 & 5.21 \\
			10 & 9.45 \\
			100 & 10.32 \\
			\hline
		\end{tabular}
		\label{cov2}
	\end{table}

	For the loading matrix $2$, B\&D had an MSE of $10.81$. Again, our method beats this across a range of values of $u$. The B\&D model is a special case of PFA with $E=I_p$ in \eqref{MFA}. We conjecture that the gains seen for PFA are due to the more flexible induced shrinkage structure on the covariance matrix. 
	%combined with avoidance of the need for post-processing to estimate the loadings matrix.  
	
	We compare the estimated and true loading matrices in Figure~\ref{nopert}. Throughout the paper, PFA estimated loadings are based on $\hat{\Lambda}\hat{E}^{1/2}$, where $\hat{\Lambda}$ and $\hat{E}$ are the estimated loading and covariance matrix of the factors, respectively. This makes the loadings comparable to methods that use identity covariance for the latent factors.
	PFA performs overwhelmingly better at estimating the true loading structure compared with B\&D FA. The first five columns of the estimated loadings based on PFA are very close to the true loading structure under some permutation. This gain over B\&D may be due to a combination of the more flexible shrinkage structure and the different post-processing scheme. On comparing $\hat{\Lambda}\hat{E}^{1/2}$ for different choices of $u$, we see that the estimates are not sensitive to the hyperparameter $u$. Estimation MSE is better for smaller $u$, but the general structure of the loadings matrix is recovered accurately for all $u$.

	\subsection{Case 2: Multi-group, multiplicative perturbation}
	\label{simu2}
	In this case, we simulate data from the multi-group model in~\eqref{GMSEM} for $i=1,\ldots,500$. Observations $Y_i$ are first generated using the same method as in case 1. The data are then split into $10$ groups of $50$ observations, such that $G_j=\{Y_k: 50(j-1)\leq k\leq 50j\}$ for $j=1,\ldots,10$. The groups are perturbed using matrices $Q_{j0}\sim$MN$(I_p, \alpha_0 I_p, \alpha_0 I_p)$ for different choices of $\alpha_0$, setting $Q_{1}= I_p$.
	
	\begin{figure}[htbp]
		\centering
		\includegraphics[width = 1\textwidth]{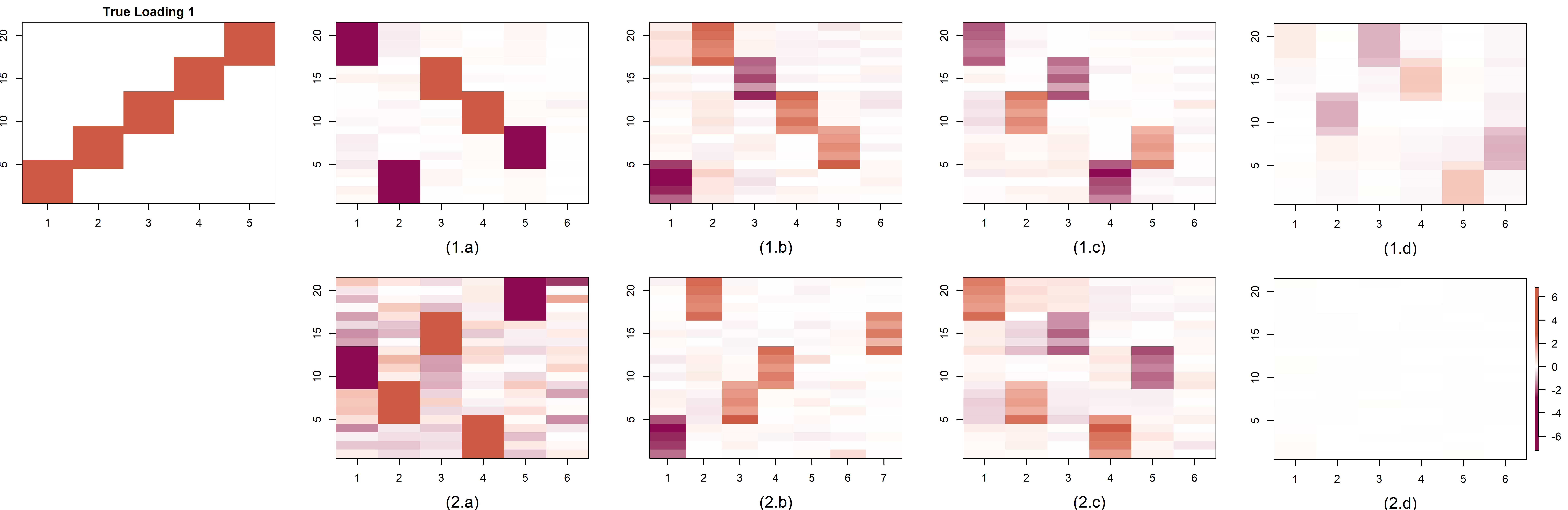}
		\caption{Comparison of estimated loading matrix 1 in simulation case 2 with different choices of $\alpha_0$ and $\alpha$ where $Q_{j0}\sim$MN$(I_p, \alpha_0 I_p, \alpha_0 I_p)$ and $U=\alpha I_p=V.$ (a) $\alpha =  10^{-4}, \alpha_0= 10^{-4}$, (b) $\alpha = 10^{-2}, \alpha_0= 10^{-4}$, (c) FBPFA with $\alpha_0= 10^{-4}$, (d) BMSFA with $\alpha_0= 10^{-4}$, (e) $\alpha =  10^{-4}, \alpha_0=10^{-2}$, (f) $\alpha = 10^{-2}, \alpha_0=10^{-2}$, (g) FBPFA with $\alpha_0=10^{-2}$, (h) For BMSFA with $\alpha_0=10^{-2}$. True loading matrices are plotted twice in columns 1 for easier comparison with other images. }
		\label{pertgrp}
	\end{figure}

	Figure~\ref{pertgrp} shows the estimated loading matrices. We obtain accurate estimates even when assuming a higher level of perturbation than the truth, $\alpha_0\leq \alpha$. However, the estimates are best when $\alpha=\alpha_0= 10^{-4}$. Performance degrades more sharply when we underestimate the level of perturbation, as in the case where $\alpha= 10^{-4}, \alpha_0=10^{-2}$. The BMSFA code requires an upper bound on the number of shared and group-specific factors; we choose both to be $6$.
	The estimated loadings from PFA and FBPFA are better than BMSFA. Except for cases with a higher level of perturbation using BMSFA, the estimated loadings matrix is close to the truth under some permutation of the columns. BMSFA estimated loadings usually are of lower magnitude, which may be due to the additive structure of the model. Thus they look faded in almost all the figures. The cumulative shrinkage prior used in PFA and BMSFA induces continuous shrinkage instead of exact sparsity, but nonetheless does a good job overall of capturing the true sparsity pattern. BMSFA estimated loadings for $\alpha_0=10^{-2}$ look blank as the estimated loading has all entries near zero. Results for the case 2 loading structure are shown in Figure 5 of the Supplementary materials.
	
	\begin{table}[htbp]
		\centering
		\caption{Average predictive log-likelihood for PFA with optimal $\alpha$, FBPFA and BMSFA in simulation case 2.}
		\begin{tabular}{r|rrrr}
			\hline
			True Loading&$\alpha_0$& PFA for optimal $\alpha$ & FBPFA&BMSFA \\
			\hline
			Loading 1 &$ 10^{-4}$ & $-$31.65 &$-25.92$& $-$679.36 \\
			&$10^{-2}$ & $-$30.32 & $-26.01$ &$-$921.76 \\
			\hline
			
			Loading 2 & $ 10^{-4}$ & $-$210.43 & $-251.13$ &$-$6871.38 \\
			& $10^{-2}$ & $-$351.42 & $-$304.10&$-$25018.50 \\
			
			\hline
		\end{tabular}
		\label{case2pll}
	\end{table}

\begin{figure}[htbp]
		\centering
		\includegraphics[width = 1\textwidth]{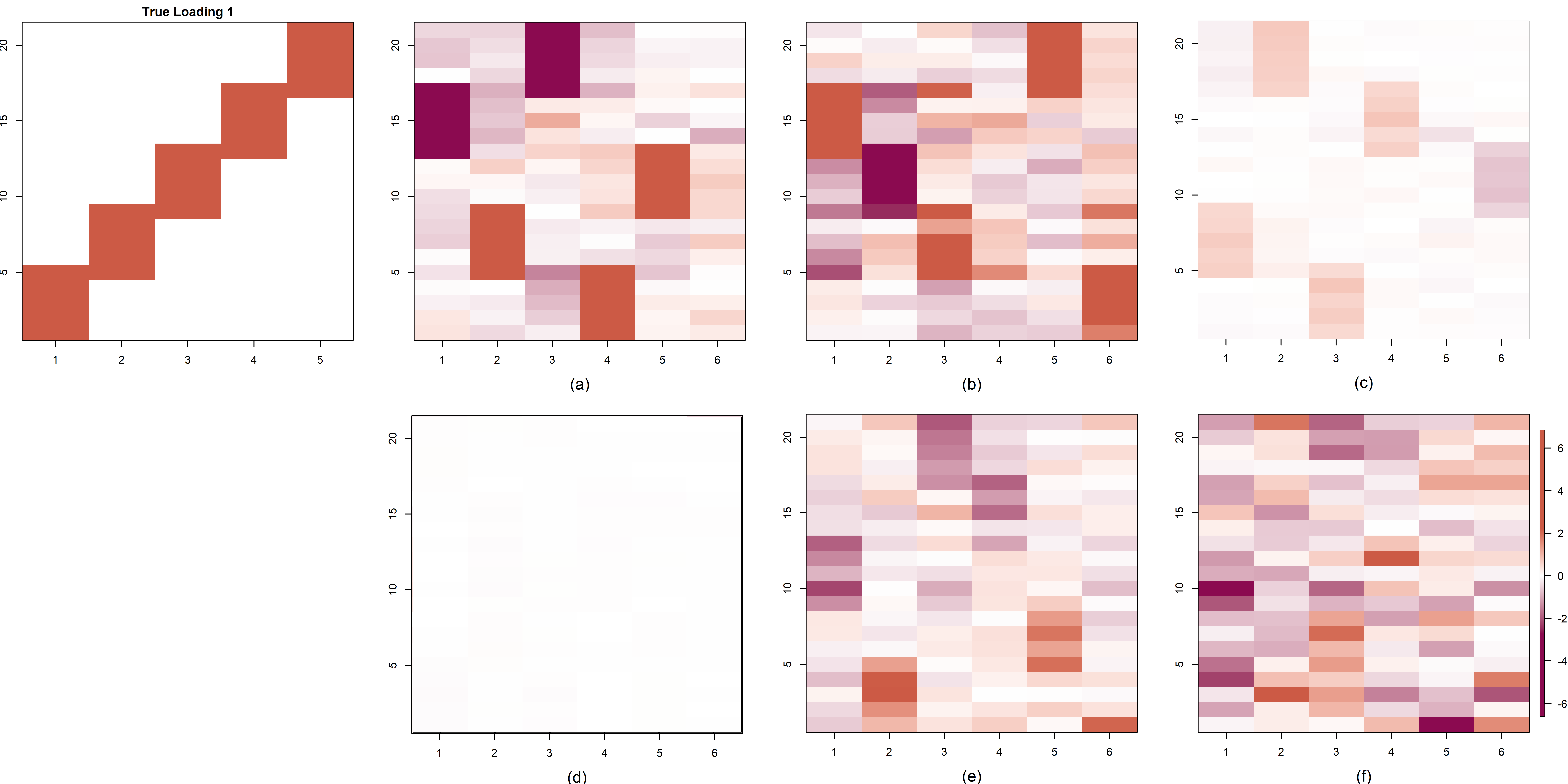}
		\caption{Comparison of estimated loading matrices in simulation case 2 with increasing error variances and $\alpha_0= 10^{-4}$: (a) $\alpha =  10^{-4}$ with error variance 25, (b) $\alpha =  10^{-4}$ with error variance 100, (c) BMSFA with error variance 25, (d) BMSFA with error variance 100, (e) FBPFA with error variance 25, (f) FBPFA with error variance 100. }
		\label{varyvar}
	\end{figure}

	We also apply the same simulation setting with higher error variances. 
	Figure~\ref{varyvar} compares the estimated loadings. %The entries in our loading matrix are of order 5 in Loading 1. 
	All the methods identify the true structure when the residual standard deviation is 5, but only PFA with correctly specified $\alpha$ produces good estimates when we increase it to 10.
	
\begin{figure}[htbp]
		\centering
		\includegraphics[width = 1\textwidth]{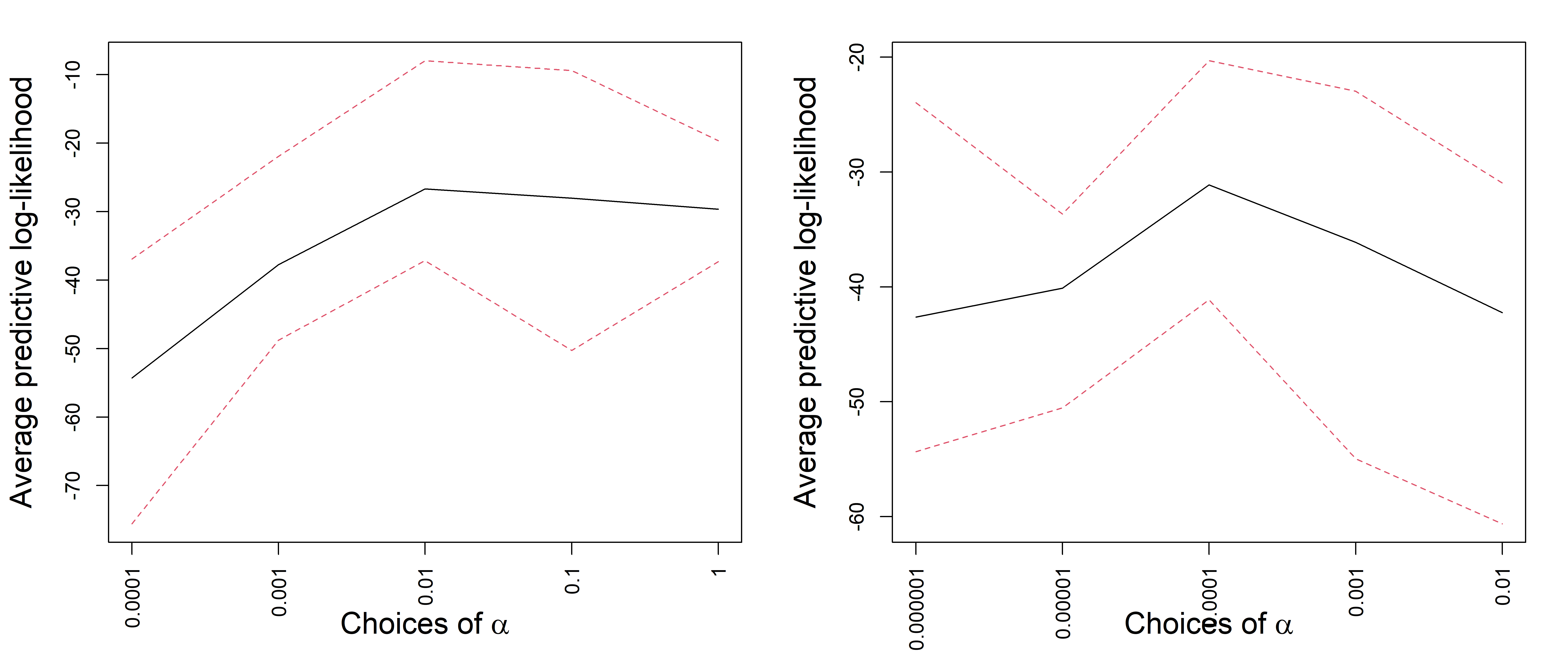}
		\caption{Average predictive log-likelihoods in black along with two dotted red lines, denoting the range of predictive log-likelihoods across all the splits for different choices of $\alpha$ in simulation case 2 for two cases - in the first case the perturbation matrices are generated using true $\alpha_0=10^{-2}$ and in the second case  $\alpha_0= 10^{-4}$. This is based on the simulation experiment of Figure~\ref{pertgrp}.}
		\label{CV}
	\end{figure}

	In Figure~\ref{CV}, we show the predictive log-likelihood averaged over all the splits under a range of $\alpha$ values. For each choice of $\alpha$, we fit the model to a training set and calculate the predictive log-likelihood on a test set for 10 randomly chosen training-test splits. This demonstrates the utility of our cross-validation technique in finding the optimal $\alpha$. In Table~\ref{case2pll}, we compare the performance of PFA with optimal $\alpha$, FBPFA, and BMSFA in terms of predictive likelihood. PFA and FBPFA have overwhelmingly better performance. We also show the range of predictive log-likelihoods for each choice of $\alpha$ in Figure~\ref{CV} using dotted red lines.
	FBPFA is better at recovering the true loading structure for $p=21$, but for the higher dimensional case with $p=128$ PFA tends to be better.
	
	\subsubsection{Case 2.1: Partially shared factors}
	We also repeat the case 2 simulation but modified to accommodate a partially shared structure.  In particular, we generated the data as in case 2 but for the last two groups we let $Q_jY_i\sim$N$(\Lambda_{01}\Lambda_{01}^T+I_p)$ for $i\in G_j$ and $j=9,10$. 
	Here $\Lambda_{01}=\Lambda_{0}[, 1:3]$ is a $p\times 3$ dimensional matrix having the first three columns from $\Lambda_0$. The choices for $\Lambda_0$ are given in Figure~\ref{true}. 
	
	\begin{figure}
		\centering
		\includegraphics[width = 1\textwidth]{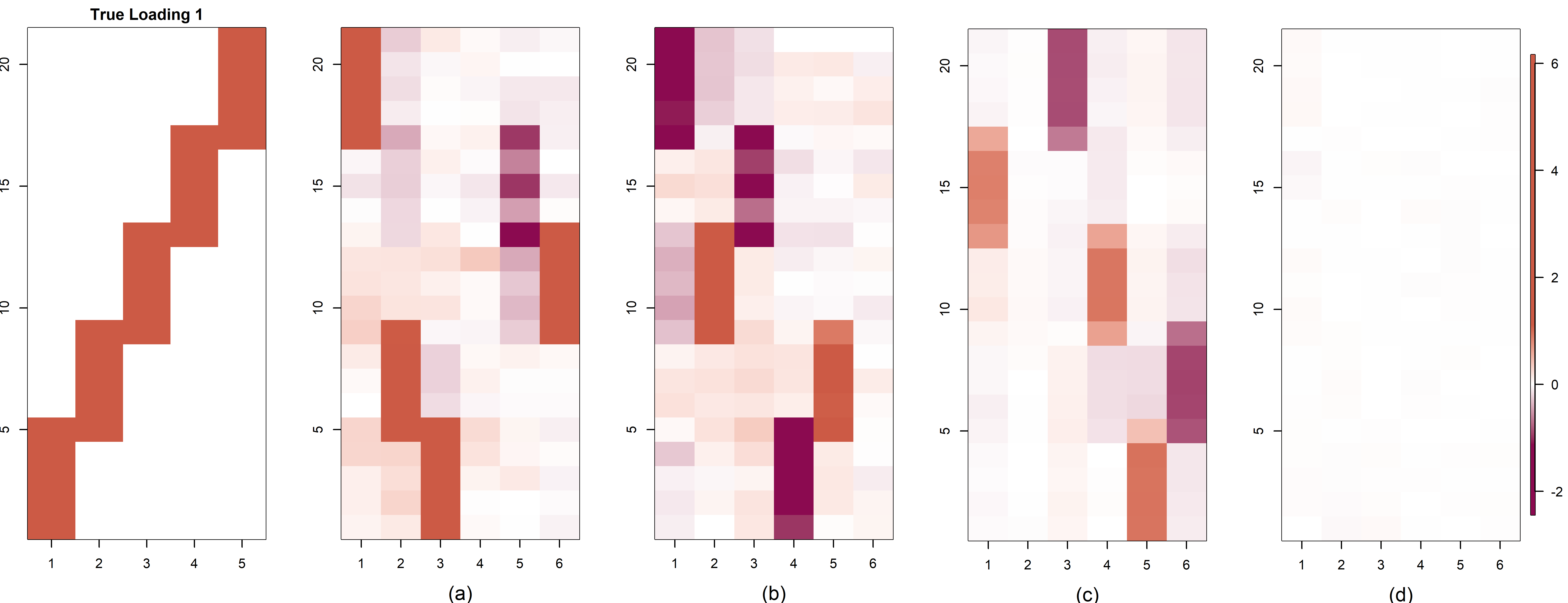}
		\caption{Comparison of estimated loading in the partially shared modification of simulation case 2.  Row 1 corresponds to true loading structure 1 and row 2 to true loading structure 2. (a) FBPFA with $\alpha_0= 10^{-4}$, (b) FBPFA with $\alpha_0=10^{-2}$, (c) BMSFA with $\alpha_0= 10^{-4}$, (d) BMSFA with $\alpha_0=10^{-2}$.}
		\label{fig:pertsome}
	\end{figure}
	
	The estimated loading matrices are compared in Figure~\ref{fig:pertsome} for two levels of perturbations, $\alpha_0= 10^{-4},10^{-2}$.  FBPFA works well for any level of perturbation, while BMSFA works well for only the lower perturbation level. The estimated loading in Figure~\ref{fig:pertsome} (d) looks blank due to near zero entries. Estimated loadings for the case 2 loading structure are in Figure 6 of the Supplementary materials.

	%The BMSFA estimated loadings are worse than PFA estimated loadings for the cases with a higher amount of perturbation when $\alpha_0=0.01$. 

	%\subsubsection{Case 2.1: Partially shared structure}
	%In this section, we repeat the 

	\subsection{Case 3: Multi-group factor model}
	In this case, we generate data from the Bayesian multi-study factor analysis (BMSFA) model as in \cite{de2018bayesian}.
	\begin{align}
		Y_i=&\Lambda\eta_{i1}+\Psi_j\eta_{i2}+\epsilon_{1i} \textrm{ and } Y_i\in G_j,\nonumber\\
		\epsilon_{1i}\sim & \textrm{N}(0, \Sigma)\quad \eta_{i1},\eta_{i2}\sim \textrm{N}(0, I_p),\label{BMSFA}
	\end{align}
	where the group-specific loadings ($\Psi_j$'s) are lower in magnitude in comparison to the shared loading matrix $\Lambda$. We generate the $\Psi_j$s from  N$(-0.5, 0.8)$.  These $\Psi_j$ matrices are the same dimension as the shared loading matrix $\Lambda$. 
	\begin{table}[htbp]
		\centering
		\caption{Average predictive log-likelihood for PFA for different choices of $\alpha$, FBPFA and BMSFA in Simulation Case 3.}
		\begin{tabular}{r|rrrrr}
			\hline
			Generative &True & PFA for & PFA for & FBPFA & BMSFA \\ 
			Distribution&Loading&$\alpha = 10^{-2}$&$\alpha =  10^{-4}$&&\\
			of $\Psi_j$'s&&&&\\
			\hline
			N$(-0.2, 0.2)$ &Loading 1&  $-$33.41 &$-$44.10 & $-$26.16 & $-$500.66 \\ 
			& Loading 2 & $-$336.58 & $-$259.79 & $-$265.77 & $-$9547.77 \\ \hline
			N$(-0.5, 0.8)$   & Loading 1 & $-$49.25& $-$53.32 & $-$37.89 & $-$720.35 \\ 
			& Loading 2 & $-$584.82& $-$308.72 & $-$1003.02 & $-$14187.36 \\ 
			\hline
		\end{tabular}
		\label{case4pll}
	\end{table}
	
\begin{figure}[htbp]
		\centering
		\includegraphics[width = 1\textwidth]{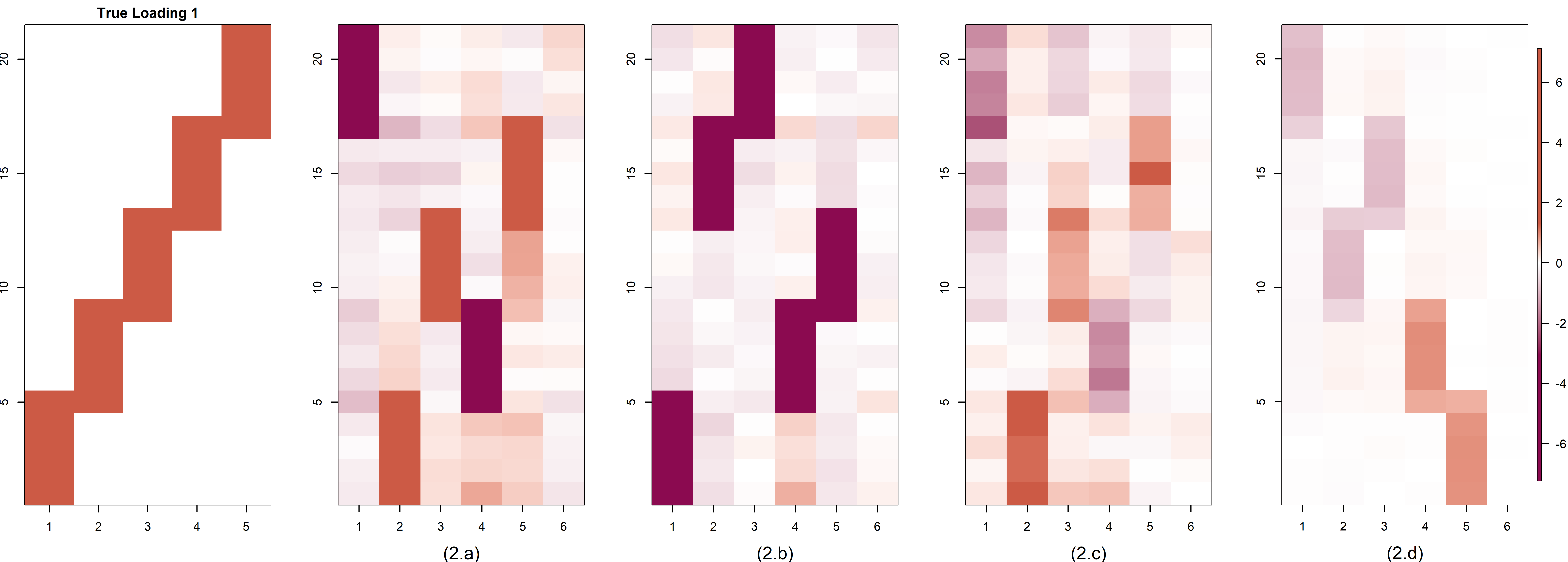}
		\caption{Comparison of estimated loading matrices in simulation case 3 when the true data generating process follows the model in~\eqref{BMSFA}. (a) PFA with $\alpha =  10^{-4}$, (b)PFA with $\alpha = 10^{-2}$, (c) FBPFA, (d) BMSFA.}
		\label{Pertnewrev}
	\end{figure}

	Figure~\ref{Pertnewrev} compares the estimated loadings with the true shared loadings across different methods: PFA for two different perturbation parameters, FBPFA and BMSFA. Although we generate the data from BMFSA, all the estimates are very much comparable in Figure~\ref{Pertnewrev}. However, PFA and FBPFA again outperform BMSFA in terms of predictive log-likelihoods as shown in Table~\ref{case4pll}. In the supplementary material, we present another simulation setting akin to BMSFA. There, the data generating process is similar to BMSFA with minor modifications in the shared and group-specific loading structures. Estimated loadings for the case 2 loading structure are in Figure 7 of the Supplementary materials.   
	
	\subsection{Case 4: Mimic NHANES data}
\begin{figure}
		\centering
		\includegraphics[width = 1\textwidth]{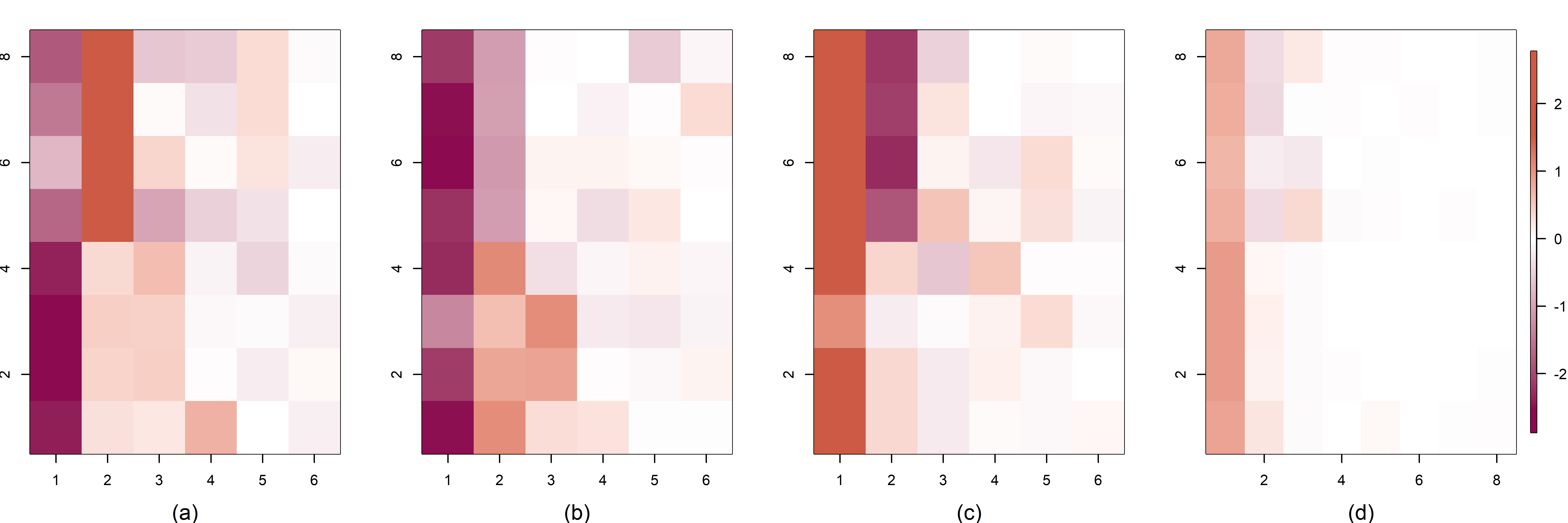}
		\caption{Comparison of estimated loading matrices in simulation case 4 where the true loading matrix is PFA estimated loading with $\alpha=10^{-2}$ for mean-centered NHANES data and the data are simulated from the model~\eqref{GMSEM}. (a) PFA with $\alpha = 10^{-2}$, (b) PFA with $\alpha =  10^{-4}$, (c) FBPFA, (d) BMSFA. The color gradient added in the last image holds for all the images.}
		\label{fig:realmimic}
	\end{figure}

	In this section, we generate data from the model in~\eqref{GMSEM}. However, the model parameters $Q_{j}$'s and $\Lambda$ are first estimated on the NHANES data with $\alpha=10^{-2}$. Then, based on these estimated parameters, we generate the data following the same model. Figure~\ref{fig:realmimic} compares the estimated loadings across all the methods: PFA for different choices of $\alpha$, FBPFA, and BMSFA. We find that all the methods recover loading structures that match with our estimated loadings from Section~\ref{real}. For PFA and FBPFA, we selected the number of factors as described in Section \ref{prior}, and hence these approaches produced fewer columns than BMSFA.

	\section{Application to NHANES data}
	\label{real}
	
	We fit our model in~\eqref{GMSEM} to the NHANES dataset as described in Section~\ref{model} to obtain phthalate exposure factors for individuals in different ethnic groups. Before analysis the mean chemical levels across all groups are subtracted to mean-center the data. The data are then randomly split, with $2/3$ in each group as a training set, and $1/3$ as a test set. We collect 5000 MCMC samples after a burn-in of 5000 samples. Convergence is monitored based on the predictive log-likelihood of the test set at each MCMC iteration. The hyperparameters are the same as those in Section~\ref{sim}. The predictive log-likelihood of the test data are used to tune $\alpha$ as described in Section~\ref{tune}. Based on our cross-validation technique, $\alpha=10^{-2}$ is chosen. Note that our final estimates are based on the complete data. Our BMSFA implementation sets the upper bounds at 8 for both shared and group-specific factors. Figure 4 of the supplementary materials depicts trace plots of the loading and perturbation matrices for FBPFA.
	
\begin{figure}[htbp]
		\centering
		\includegraphics[width = 1\textwidth]{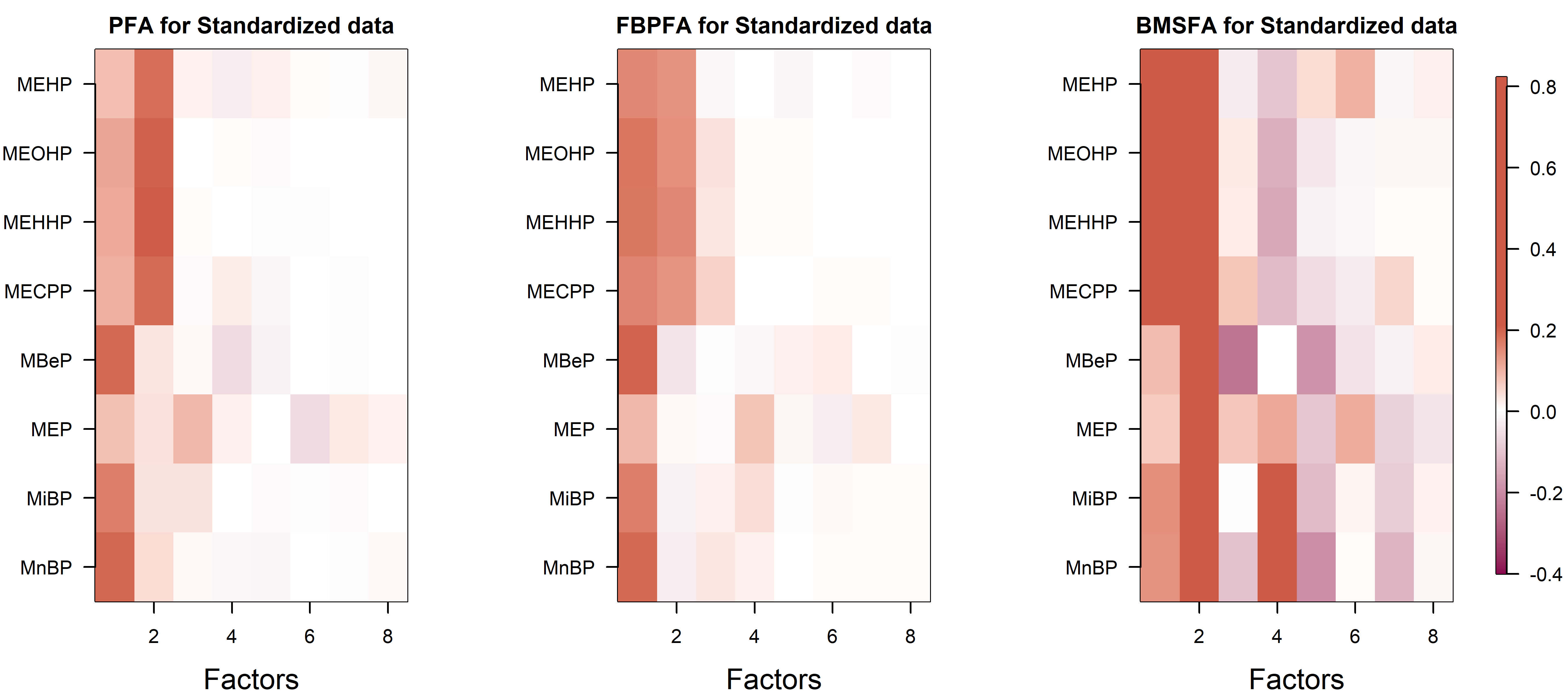}
		\caption{Comparison of estimated loading matrices with different methods. PFA estimates are based on $\alpha=10^{-2}$.}
		\label{loadreal}
	\end{figure}

	Figure~\ref{loadreal} shows the estimated loadings using PFA, FBPFA, and BMSFA. All plots show two significant factors, with some suggestions of a third factor. For PFA and FBPFA, all the chemicals load on the first factor, and the second factor is loaded on by the first four chemicals, namely MEHP, MEOHP, MEHHP, and MECPP. 
	Figure~\ref{realcov} also suggests that these four chemicals are related to each other more than the others. This is not surprising, as MECPP, MEHHP, and MEOHP are oxidative metabolities of MEHP. 
	
	Comparing the predictive log-likelihoods for PFA, FBPFA and BMSFA, we obtain values of $4.12$, $5.02$ and $-7.99$, respectively, suggesting much better performance for the PFA-based approaches.  As an alternative measure of predictive performance, we also consider test sample MSE using 2/3 of the data as training and 1/3 as test.  The predictive value of the test data is estimated by averaging the predictive mean conditionally on the parameters and training data over samples generated from the training data posterior.  We obtain predictive MSE values of $0.05$ and $0.09$ for PFA with $\alpha= 10^{-4}$ and $\alpha=10^{-2}$, respectively.  FBPFA yields a value of $0.14$, while BMSFA has a much larger error of $0.70$.  Considering that the data are normalized before analysis, the naive prediction that sets all the values to zero would yield an MSE of $1.0.$

\begin{figure}[htbp]
		\centering
		\includegraphics[width = 0.4\textwidth]{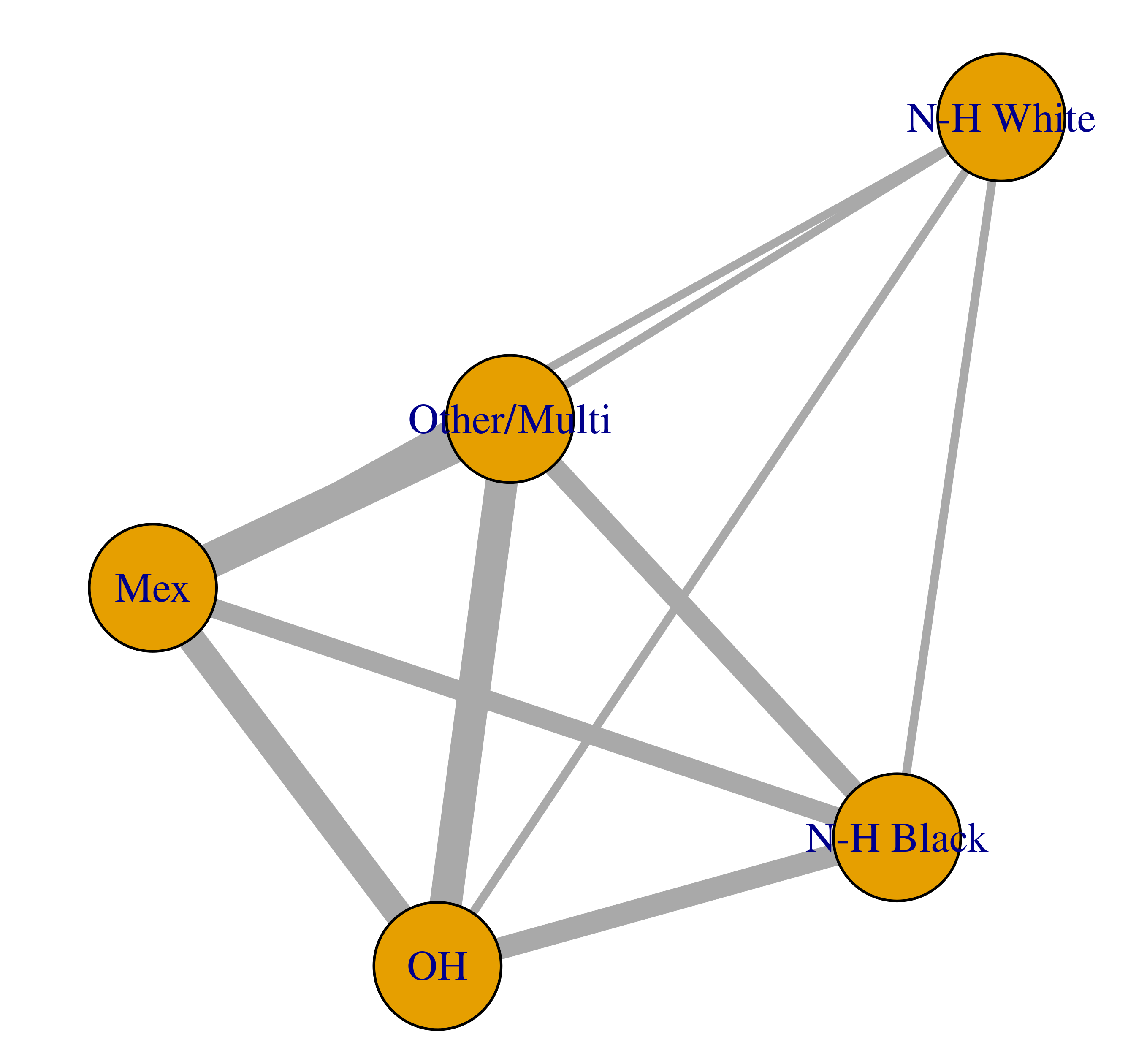}
		\caption{Network plot summarizing similarity between ethnic groups based on the divergence metric with thicker edges implying smaller divergence scores. Nodes correspond to different ethnic groups.}
		\label{realnet}
	\end{figure}

	To explore similarities across groups, 
	we calculate divergence scores and create the divergence matrix $D$. This matrix is provided in Table~\ref{realtable}. To summarize $D$, we also provide a network plot between different groups in Figure~\ref{realnet}.  The plot uses greater edge widths between ethnic groups when the divergence statistic between the groups is small.  In particular, the edge width between two nodes $(j, l)$ is calculated as $60/d_{jl}$.  Thicker edges imply greater similarity. This figure implies that non-Hispanic whites differ the most in phthalate exposure profiles from the other groups, supporting the findings of our exploratory ANOVA analysis in the Section 3 of the supplementary materials and findings in the prior literature \citep{bloom2019racial}. We find evidence of one extra factor among non-Hispanic whites relative to the other groups in Figure~\ref{realCF}. The two Hispanic ethnicity groups are very similar in loadings as shown in Figure~\ref{realCF} and also their Hotelling $T^2$ distance is relatively small in Table~\ref{htsqr}.  %Our results based on divergence scores also suggest similar distinctions among the groups. 
	These results support our preliminary analysis, described in Section~\ref{model}. 
	
	In Table~\ref{divBMSFA}, we show square norm differences between the group-specific loadings, estimated using BMSFA.% and the associated post-processing {\bf [DD - what do you mean the associated post-processing?? Presumably all the BMSFA estimated loadings we show are after post-processing]}. 
	The $(i, j)$-$th$ entry of the table is calculated as $\sqrt{\sum_{lk}(A_{ill}-A_{jlk})^2}$, where the matrices $A_i$ and $A_j$ are the estimated group-specific loadings for the groups $i$ and $j$, respectively, using BMSFA.  Based on these results, BMSFA-based inferences on differences in exposure profiles across ethnic groups are noticeably different from our PFA-based results reported above. In particular, all the groups seem to be approximately the same distance apart. {The results using BMSFA from Table~\ref{divBMSFA} are not consistent with Hotelling $T^2$-based our exploratory data analysis results from Table~\ref{htsqr} and PFA-based results from Table~\ref{realtable}.} 
	%Figure~\ref{loadrealnonstan} shows a similar kind of latent structure in the non-standardized data.

	\begin{table}[ht]
		\centering
		\caption{Estimated divergence scores for different pairs of ethnic groups based on NHANES data.}
		\begin{tabular}{rrrrrr}
			\hline
			& Mex & OH & N-H White & N-H Black & Other/Multi \\ 
			\hline
			Mex & 0.00 & 5.86 & 41.35 & 10.22 & 3.82 \\ 
			OH & 5.86 & 0.00 & 40.93 & 8.38 & 4.45 \\ 
			N-H White & 41.35 & 40.93 & 0.00 & 40.06 & 41.17 \\ 
			N-H Black & 10.22 & 8.38 & 40.06 & 0.00 & 9.48 \\ 
			Other/Multi & 3.82 & 4.45 & 41.17 & 9.48 & 0.00 \\ 
			\hline
		\end{tabular}
		\label{realtable}
	\end{table}

	\begin{table}[ht]
		\centering
		\caption{Square norm difference of the group-specific loadings, estimated using BMSFA.}
		\begin{tabular}{rrrrrr}
			\hline
			& Mex & OH & N-H White & N-H Black & Other/Multi \\ 
			\hline
			Mex & 0.00 & 0.75 & 0.70 & 0.59 & 0.78 \\ 
			OH & 0.75 & 0.00 & 0.63 & 0.75 & 0.73 \\ 
			N-H White & 0.70 & 0.63 & 0.00 & 0.68 & 0.84 \\ 
			N-H Black & 0.59 & 0.75 & 0.68 & 0.00 & 0.91 \\ 
			Other/Multi & 0.78 & 0.73 & 0.84 & 0.91 & 0.00 \\ 
			\hline
		\end{tabular}
		\label{divBMSFA}
	\end{table}
	
	\section{Discussion}
	\label{discuss}
	
	In this paper, our focus has been on identifying a common set of phthalate exposure factors that hold across different ehtnic groups, while also inferring differences in exposure profiles across groups. To accomplish this goal, we focused on data from NHANES, which contain rich information on ethnicity and exposures.  We found that our proposed perturbed factor analysis (PFA) approach had significant advantages over existing approaches in addressing our goals.  
	
	There are multiple important next steps to consider to expand on our analysis.  A first is to include covariates.  In addition to phthalates and ethnic group data, NHANES collects information that may be relevant to understanding ethnic differences in exposure profiles, including BMI, age, socioeconomic status, gender and other factors.  
	\cite{li2017incorporating} consider a related problem of incorporating covariates into factor models for multiview data.  Perhaps the simplest and most interpretable modification of our PFA model of the NHANES data is to allow the factor scores $\eta_{ij}$ to depend on covariates $X_{ij}=(1,X_{ij2},\ldots,X_{ijq})^T$ through a latent factor regression.
	For example, we could let $\eta_{ij} = \beta X_{ij} + \xi_{ij}$, with $\beta$ a $k \times q$ matrix of coefficients and $\xi_{ij} \sim \textrm{N}(0,E).$  
	
	Another important direction is to extend the analysis to allow inferences on relationships between exposure profiles and health outcomes.  NHANES contains rich data on a variety of outcomes that may be adversely affected by phthalate exposures.  To include these outcomes in our analysis, and effectively extend PFA to a supervised context, we can define separate PFA models for the exposures and outcomes, with these models having shared factors $\eta_{ij}$ but different perturbation matrices.  It is straightforward to modify the Gibbs sampler used in our analyses to this case and/or the extension described above to accommodate covariates $X_{ij}$; further modifications to mixed continuous and categorical variables can proceed as in \cite{carvalho2008high, zhou2015bayesian} with the perturbations conducted on underlying variables.  Relying on such a broad modeling and computational framework, it would be interesting to attempt to infer causal relationships between ethnicity and phthalates and adverse health outcomes.  Mediation analysis may provide a useful framework in this respect.
	
	Beyond this application, the proposed PFA approach may prove useful in other contexts.  The important features of PFA include both the incorporation of a multiplicative perturbation of the data and the use of heteroscedastic latent factors.  These innovations are potentially useful beyond the multiple group setting to meta analysis, measurement error models, nonlinear factor models including Gaussian process latent variable models \citep{lawrence2004gaussian,lawrence2006local} and variational auto-encoders \citep{kingma2013auto,pu2016variational}, and even to obtaining improved performance in `vanilla' factor modeling lacking hierarchical structure.  
	
	Code is available for implementing the proposed approach and replicating our results from Section~\ref{simu2} at {\url {https://github.com/royarkaprava/Perturbed-factor-model}}.

	%For simplicity, we have considered identical entries in the diagonal of $U$ and $V$. This suggests homoscedastic perturbation across all the variables. However, one potential future direction is to allow for different levels of perturbations across different variables. Thus one can assume $U=$diag$(\alpha_1,\ldots,\alpha_p)$ and similarly on $V$. If inter-group variations are large, one possibility is to consider group-specific perturbation parameters such as $Q_j\sim$MN$(I_p,\alpha_jI_p,\alpha_jI_p)$. Since it requires to choose optimal $\alpha_j$'s simultaneously for all the groups, considering the FBPFA framework is more suitable. Thus, we can put weakly informative IG$(0.1,0.1)$ prior of $\alpha_j$. Furthermore, the shape parameter $u$ is kept fixed at 100. In Figure~\ref{nopert}, we find that the estimates are not very sensitive to the choice of $u$ in recovering true loading structure. However, one can put a hyper prior on $u$ such as weaky informative IG$(0.1,0.1)$ and make more robust inferences.
	
	\section{Acknowledgments}
	This research was partially supported by
	grant R01-ES027498 from the National Institute of Environmental Health Sciences (NIEHS) of the National
	Institutes of Health (NIH). We would like to thank Roberta De Vito for sharing her source code of the Bayesian MSFA model. We would also like to thank Noirrit Kiran Chandra for his feedback on the code which heavily improved its usage both in low and high-dimensional settings.

	\section{Supplementary Materials}
	
	\subsection{Posterior consistency}
	\label{GPOST}
	
	Let $\Theta_{\Lambda}, \Theta_{\Sigma}, \Theta_{E}$ be the parameter spaces of $\Lambda, \Sigma$ and $E$, respectively, $\Theta$ be the set of $p\times p$ positive semidefinite matrices corresponding to the parameter space of $H$, and $\mathcal{Q}_j$ be the parameter space of $Q_j$. Let $\Pi_{\Lambda},\Pi_{\Sigma}, \Pi_{E},\Pi_{Q}$ be the priors for $\Lambda, \Sigma,E$ and $Q_j$'s. We restate some of the results from \cite{bhattacharya2011sparse} for our modified factor model. With minor modification, the proofs will remain the same.
	
	Let $g:\Theta_{\Lambda}\times\Theta_{\Sigma}\times \Theta_{E}\rightarrow \Theta_{H}$ be a continuous map such that $g(\Lambda, \Sigma,E)=\Lambda^{T}E\Lambda+\Sigma$.
	
	\begin{lemma}
		For any $(\Lambda, \Sigma, E)\in \Theta_{\Lambda}\times\Theta_{\Sigma}\times \Theta_{E}$, we have $g(\Lambda, \Sigma,E)\in \Theta_{H}$.
	\end{lemma}
	
	The proof is similar to Lemma 1 of \cite{bhattacharya2011sparse}. In our Bayesian approach, we choose independent priors for $\Lambda$, $\Sigma$ and $E$ and that induces a prior on $H$ through the map $g$. We also have following proposition.
	
	\begin{proposition}
		If $(\Lambda, \Sigma, E)\sim \Pi_{\Lambda}\otimes\Pi_{\Sigma}\otimes\Pi_{E}$, then $\Pi_{\Lambda}\otimes\Pi_{\Sigma}\otimes\Pi_{E}(\Theta_{\Lambda}\times\Theta_{\Sigma}\times \Theta_{E})=1$.
	\end{proposition}
	The proof is similar to Proposition 1 of \cite{bhattacharya2011sparse} with minor modifications. Now, we proceed to establish that the posterior of our multigroup model is weakly consistent under a fixed $p$ and increasing $n$ regime. Let us assume that the complete parameter space is $\kappa=(\Lambda, \Sigma,E,Q_2,\ldots, Q_j)$ and let $\kappa_0$ be the truth for $\kappa$.\\ %We make following assumptions on $\kappa_0$.
	
	\noindent
	{\it Assumptions:}
	\begin{enumerate}
		\item For some $M>0$, the true perturbation matrices are $Q_{j0}\in \mathcal{C}_{M}$, with $\mathcal{C}_{M}$ defined in Section 2.1.1.
		\item There exists some $E>0$, $F_1>0$ and $F_2>0$ such that $\max_{ij}|\Lambda_0|<E$ and $F_1<e_{0k}<F_2$ for all $k=1,\ldots r$.
	\end{enumerate}
	
	\begin{theorem}
		\label{contheo}
		Under Assumptions 1-2, the posterior for $\kappa$ is weakly consistent at $\kappa_0$.
	\end{theorem}
	
	We first show that our proposed prior has large support in the sense that the truth belongs to the Kullback-Leibler support of the prior. Thus the posterior probability of any neighbourhood around the truth converges to one in $P_{\kappa_0}^{(n)}$-probability as $n$ goes to $\infty$ as a consequence of \cite{schwartz1965bayes}. Here $P_{\kappa}^{(n)}$ is the distribution of a sample of $n$ observations with parameter $\kappa$. Hence, the posterior is weakly consistent.
	
	\begin{proof}
		For $q, q^* \in$ the space of probability measure $ \mathcal{P}$, let the Kullback-Leibler divergences be given by 
		$$
		KL(q^*, q) = \int q^*\log{\frac{q^*}{q}}.
		$$
		Let $K(\kappa_0, \kappa)$ denote the Kullback-Leibler divergence $$\sum_{j=1}^JKL(\textrm{N}(0, Q_j^{-1}[\Lambda E\Lambda^T+\Sigma](Q_j^{T})^{-1}), \textrm{N}(0, Q_{j0}^{-1}[\Lambda_0 E_{0}\Lambda_0^T+\Sigma_{0}](Q_{j0}^{-1})^T)).$$ For our model, we have
		\begin{align*}
			K(\kappa_0, \kappa) &= \frac{1}{2n}\Bigg[\sum_{j=1}^J - n_j\log |Q_{j0}^{-1}H_0(Q_{j0}^T)^{-1}\{Q_{j}^{-1}H (Q_{j}^T)^{-1}\}^{-1}| \\&\quad + n_j\textrm{tr}\big(Q_{j0}^{-1}H_0(Q_{j0}^T)^{-1}\{Q_{j}^{-1}H (Q_{j}^T)^{-1}\}^{-1}-I_p\big)\Bigg]
		\end{align*}
		To prove Theorem~\ref{contheo}, we rely on the following Lemma.
		
		\begin{lemma}
			\label{lem1}
			For any $\epsilon>0$, there exists $\epsilon_1>0,\epsilon_2>0$, and $\epsilon_{3j}>0$ for $j\in\{2,\ldots,J\}$ such that $\|\Lambda-\Lambda_0\|_F^2\leq \epsilon_1^2$, $\|E-E_{0}\|_F^2\leq \epsilon_2^2$ and $\|Q_j-Q_{j0}\|_F^2\leq \epsilon_{3j}^2$, then we have
			\begin{align*}
				&\Pi\big\{K(\kappa_0,\kappa)\leq \epsilon\big\} \geq \\&\quad\Pi\big\{\|\Lambda-\Lambda_0\|_F^2\leq \epsilon_1^2, \|E-E_{0}\|_F^2\leq \epsilon_2^2, \|Q_j-Q_{j0}\|_F^2\leq \epsilon_{3j}^2, j=2,\ldots, J\big\}
			\end{align*}
			
		\end{lemma}
		%We rewrite tr$Q_{j0}^T\Sigma_0Q_{j0}\{Q_{j}^T\Sigma Q_{j}\}^{-1}$=tr$(A_j^T\Sigma_0A_j\Sigma^{-1})$, where $A_j=Q_{j}^{-1}Q_{j0}$. We use the inequality 2tr$(AB)\leq \textrm{tr}(A^2)+\textrm{tr}(B^2).$ and the result $\textrm{tr}(AB)=\textrm{tr}(BA)$. 
		Due to continuity of the functions such as determinant, trace and $g(\cdot)$, the above result is immediate following the proof of Theorem 2 in \cite{bhattacharya2011sparse}. For our proposed priors, the prior probability of the R.H.S. of Lemma~\ref{lem1} is positive. Thus the prior probability of any Kullback-Leibler neighborhood around the truth is positive. This proves Theorem~\ref{contheo}.
	\end{proof}
	
	\subsection{Additional simulations}
	
	\subsubsection{Case 1: Multi-group, additive perturbation}
	\label{supcase1}
	
\begin{figure}[htbp]
		\centering
		\includegraphics[width = 1\textwidth]{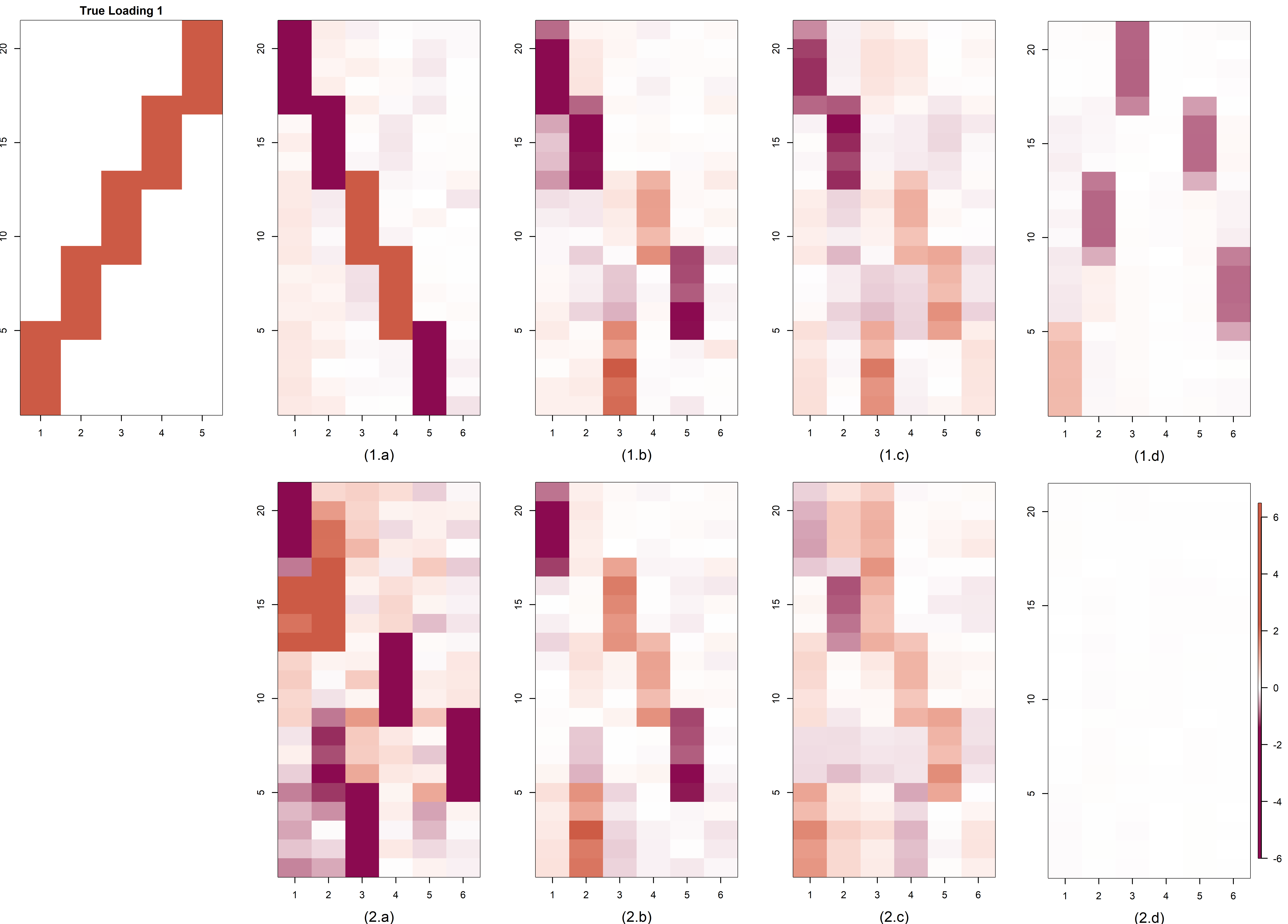}
		\caption{Comparison of estimated loading matrices in Simulation case~\ref{supcase1} when the true data generating process follows the model in~\eqref{LPert}. For the first two rows, true $\Psi_j$'s are generated from N$(-0.2, 0.2)$ and for the last two rows $\Psi_j$'s are generated from N$(-0.5, 0.8)$ and for each row (a) PFA with $\alpha = 10^{-4}$, (b) PFA with $\alpha =  10^{-2}$, (c) FBPFA, (d) BMSFA}
		\label{Pertnew}
	\end{figure}

	In Section 5.2 of the manuscript, each group is multiplied by a unique perturbation matrix $Q_j$. In this case, we perturb the data by adding a group-specific loading matrix $\Psi_j$ to a shared loading matrix $\Lambda$, as in model~\eqref{LPert}. 
	\begin{align}
		Y_i=&(\Lambda+\Psi_j)\eta_{i}+\epsilon_{1i} \textrm{ and } Y_i\in G_j,\nonumber\\
		\epsilon_{1i}\sim & \textrm{N}(0, \Sigma)\quad \eta_{i}\sim \textrm{N}(0, I_p),\label{LPert}
	\end{align}
	where the group-specific loadings ($\Psi_j$'s) are lower in magnitude in comparison to the shared loading matrix $\Lambda$. This can also be an alternative perturbation model. The group-specific loadings are generated in Figure~\ref{Pertnew} which shows the true and estimated loading matrices from BMFSA, FBPFA, and PFA across a range of values of $\alpha$. The estimated loadings from PFA are much closer to the truth than BMFSA. We find that the estimated loading matrices are some permutation of the true loading matrix with some exceptions. For the case with higher perturbation (row 2 of Figure~\ref{Pertnew}), BMSFA estimate is not good and PFA estimate is also bad for lower values of $\alpha$. Similarly, for the case with lower perturbations, PFA estimate with lower $\alpha$ is better than higher $\alpha$ estimated loading. Table~\ref{case3pll} compares predictive likelihoods of the two methods in different cases, and again, PFA and FBPFA outperform BMSFA.
	
	\begin{table}[htbp]
		\centering
		\caption{Average predictive log-likelihood for PFA for different choices of $\alpha$, FBPFA and BMSFA in Simulation Case 1.}
		\begin{tabular}{r|rrrrr}
			\hline
			Generative &True& PFA for & PFA for & FBPFA & BMSFA \\ 
			Distribution of $\Psi_j$'s&Loading&$\alpha = 10^{-2}$&$\alpha =  10^{-4}$&&\\
			\hline
			N$(-0.2, 0.2)$ &Loading 1& $-56.27$ & $-60.71$ & $-56.63$ & $-$429.28 \\ 
			& Loading 2& $-1041.28$ & $-361.82$ & $-421.28$ & $-$5853.65 \\ 
			\hline
			N$(-0.5, 0.8)$   & Loading 1& $-$30.68 & $-$50.09 & $-54.62$ & $-$514.67 \\ 
			& Loading 2& $-$320.04 & $-$246.06 & $-479.59$ & $-$6798.37 \\ 
			\hline
		\end{tabular}
		\label{case3pll}
	\end{table}
	
	\subsubsection{Case 2: Observation-level perturbations}
	\label{supcase2}
	
	In this case, we generate data from the model (2.2). First, a non-perturbed dataset is generated exactly the same way as in Section 5.1 and 5.2 in the manuscript. Then, we generate separate perturbation matrices $Q_{i0} \sim\text{MN}(I_p, \alpha_0 I_p, \alpha_0 I_p)$ for each data vector $Y_i$. We repeat this simulation for different choices of $\alpha_0$.

	\begin{figure}[htbp]
		\centering
		\includegraphics[width = 0.8\textwidth]{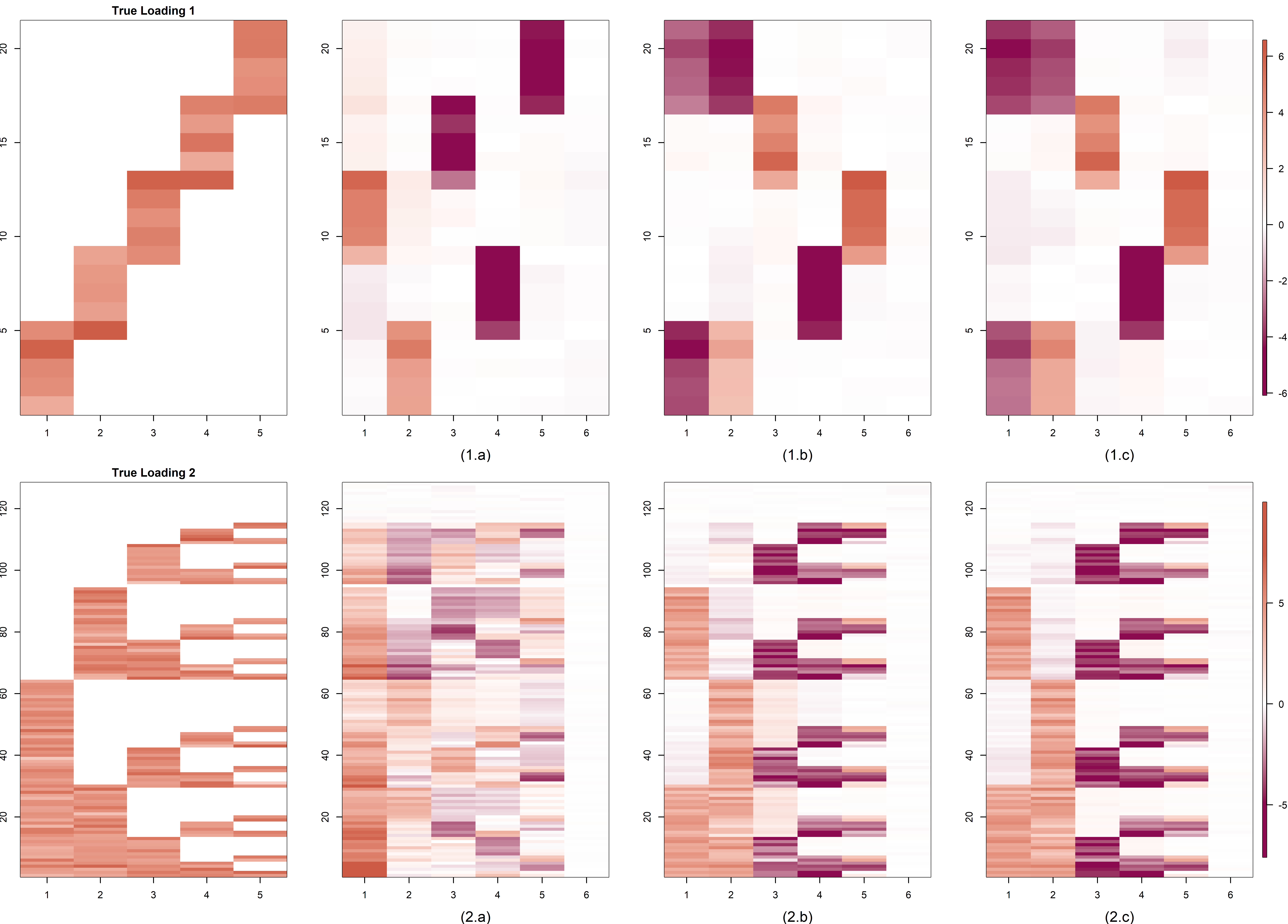}
		\caption{Estimated loadings using PFA for different choices of $\alpha$ in Simulation case~\ref{supcase2} for $\alpha_0= 10^{-4}$ and two loadings. (a): Estimated loading for $\alpha=10^{-2}$, (b): For $\alpha= 10^{-4}$, (c): For $\alpha= 10^{-5}$.}
		\label{Pertcompareall}
	\end{figure}
	
	\begin{figure}[htbp]
		\centering
		\includegraphics[width = 0.8\textwidth]{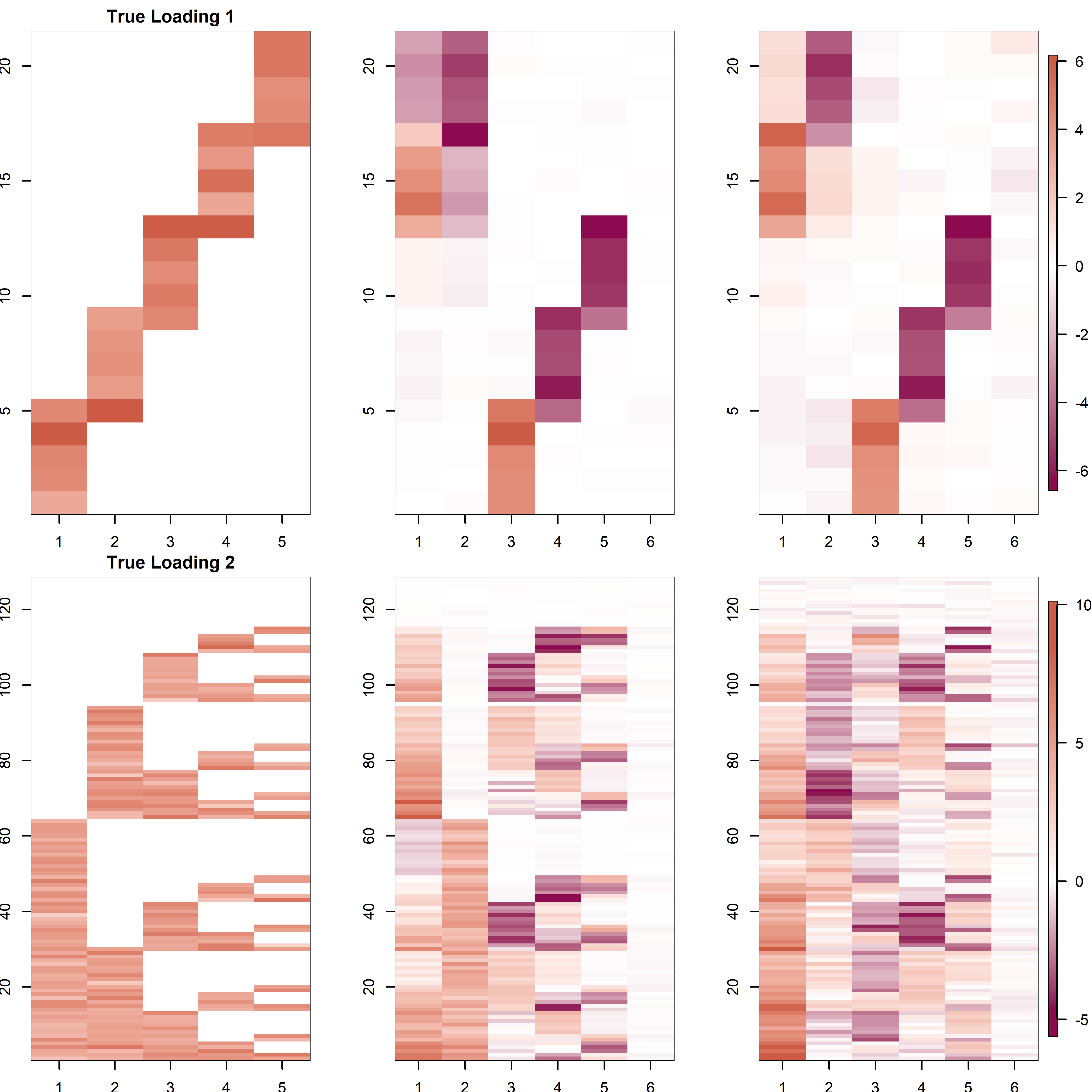}
		\caption{Estimated loadings using FBPFA for different choices of $\alpha_0$ in Simulation case~\ref{supcase2} for two cases and two loadings. (a): Estimated loading for $\alpha_0= 10^{-4}$, (b): For $\alpha_0=10^{-2}$.}
		\label{FBPFAcase4}
	\end{figure}
	
	Figures~\ref{Pertcompareall} shows the estimated loading  when they are estimated using $\alpha= 10^{-4}$. The method performs poorly when the perturbation parameter $\alpha$ is lower than the true parameter $\alpha_0$ as in the previous cases. Figure~\ref{FBPFAcase4} illustrates the performance of FBPFA in this case. When the perturbation is higher, the performance deteriorates as it is more difficult to capture the loading structure accurately. Overall, our method is able to accurately estimate the true loading structure under some permutation of the columns.

	\subsubsection{More exploratory analysis on NHANES data}
	\label{exponhanes}
	
	For each chemical level, we fit an one-way ANOVA model to analyse group-specific effects on each phthalate level separately. The results are provided from Table~\ref{first} to \ref{last}.
	
	\begin{table}[ht]
		\centering
		\caption{Estimated group effects from a one-way ANOVA analysis for the phthalate MnBP.} 
		\begin{tabular}{rrrrr}
			\hline
			& Estimate & Std. Error & t value & Pr($>|t|$) \\ 
			\hline
			(Intercept) & 3.85 & 0.11 & 36.27 & 0.00 \\ 
			N-H Black & 0.36 & 0.15 & 2.33 & 0.02 \\ 
			N-H White & -0.29 & 0.13 & -2.26 & 0.02 \\ 
			OH & 0.46 & 0.18 & 2.55 & 0.01 \\ 
			Other/Multi & -0.09 & 0.22 & -0.41 & 0.68 \\ 
			\hline
		\end{tabular}
		\label{first}
	\end{table}
	% latex table generated in R 3.5.1 by xtable 1.8-3 package
	% Thu Oct  3 13:15:25 2019
	\begin{table}[ht]
		\centering
		\caption{Estimated group effects from a one-way ANOVA analysis for the phthalate MiBP.} 
		\begin{tabular}{rrrrr}
			\hline
			& Estimate & Std. Error & t value & Pr($>|t|$) \\ 
			\hline
			(Intercept) & 2.88 & 0.06 & 49.23 & 0.00 \\ 
			N-H Black & 0.59 & 0.08 & 6.97 & 0.00 \\ 
			N-H White & -0.35 & 0.07 & -4.98 & 0.00 \\ 
			OH & 0.33 & 0.10 & 3.28 & 0.00 \\ 
			Other/Multi & -0.08 & 0.12 & -0.68 & 0.50 \\ 
			\hline
		\end{tabular}
		
	\end{table}
	% latex table generated in R 3.5.1 by xtable 1.8-3 package
	% Thu Oct  3 13:15:25 2019
	\begin{table}[ht]
		\centering
		\caption{Estimated group effects from a one-way ANOVA analysis for the phthalate MEP.} 
		\begin{tabular}{rrrrr}
			\hline
			& Estimate & Std. Error & t value & Pr($>|t|$) \\ 
			\hline
			(Intercept) & 8.32 & 0.29 & 28.80 & 0.00 \\ 
			N-H Black & 2.84 & 0.42 & 6.79 & 0.00 \\ 
			N-H White & -1.41 & 0.35 & -4.02 & 0.00 \\ 
			OH & 1.18 & 0.49 & 2.40 & 0.02 \\ 
			Other/Multi & -0.99 & 0.60 & -1.64 & 0.10 \\ 
			\hline
		\end{tabular}
		
	\end{table}
	% latex table generated in R 3.5.1 by xtable 1.8-3 package
	% Thu Oct  3 13:15:25 2019
	\begin{table}[ht]
		\centering
		\caption{Estimated group effects from a one-way ANOVA analysis for the phthalate MBeP.} 
		\begin{tabular}{rrrrr}
			\hline
			& Estimate & Std. Error & t value & Pr($>|t|$) \\ 
			\hline
			(Intercept) & 2.79 & 0.07 & 39.90 & 0.00 \\ 
			N-H Black & 0.27 & 0.10 & 2.66 & 0.01 \\ 
			N-H White & -0.09 & 0.08 & -1.05 & 0.29 \\ 
			OH & -0.01 & 0.12 & -0.06 & 0.95 \\ 
			Other/Multi & -0.23 & 0.15 & -1.57 & 0.12 \\ 
			\hline
		\end{tabular}
		
	\end{table}
	% latex table generated in R 3.5.1 by xtable 1.8-3 package
	% Thu Oct  3 13:15:25 2019
	\begin{table}[ht]
		\centering
		\caption{Estimated group effects from a one-way ANOVA analysis for the phthalate MECPP.} 
		\begin{tabular}{rrrrr}
			\hline
			& Estimate & Std. Error & t value & Pr($>|t|$) \\ 
			\hline
			(Intercept) & 4.80 & 0.12 & 41.52 & 0.00 \\ 
			N-H Black & -0.44 & 0.17 & -2.63 & 0.01 \\ 
			N-H White & -0.64 & 0.14 & -4.56 & 0.00 \\ 
			OH & -0.25 & 0.20 & -1.28 & 0.20 \\ 
			Other/Multi & -0.46 & 0.24 & -1.89 & 0.06 \\ 
			\hline
		\end{tabular}
		
	\end{table}
	% latex table generated in R 3.5.1 by xtable 1.8-3 package
	% Thu Oct  3 13:15:25 2019
	\begin{table}[ht]
		\centering
		\caption{Estimated group effects from a one-way ANOVA analysis for the phthalate MEHHP.} 
		\begin{tabular}{rrrrr}
			\hline
			& Estimate & Std. Error & t value & Pr($>|t|$) \\ 
			\hline
			(Intercept) & 3.83 & 0.10 & 36.85 & 0.00 \\ 
			N-H Black & -0.04 & 0.15 & -0.26 & 0.79 \\ 
			N-H White & -0.39 & 0.13 & -3.06 & 0.00 \\ 
			OH & -0.10 & 0.18 & -0.56 & 0.58 \\ 
			Other/Multi & -0.26 & 0.22 & -1.19 & 0.23 \\ 
			\hline
		\end{tabular}
		
	\end{table}
	% latex table generated in R 3.5.1 by xtable 1.8-3 package
	% Thu Oct  3 13:15:25 2019
	\begin{table}[ht]
		\centering
		\caption{Estimated group effects from a one-way ANOVA analysis for the phthalate MEOHP.} 
		\begin{tabular}{rrrrr}
			\hline
			& Estimate & Std. Error & t value & Pr($>|t|$) \\ 
			\hline
			(Intercept) & 3.11 & 0.08 & 39.04 & 0.00 \\ 
			N-H Black & -0.06 & 0.12 & -0.54 & 0.59 \\ 
			N-H White & -0.33 & 0.10 & -3.39 & 0.00 \\ 
			OH & -0.11 & 0.14 & -0.83 & 0.41 \\ 
			Other/Multi & -0.26 & 0.17 & -1.54 & 0.12 \\ 
			\hline
		\end{tabular}
		
	\end{table}
	% latex table generated in R 3.5.1 by xtable 1.8-3 package
	% Thu Oct  3 13:15:25 2019
	\begin{table}[ht]
		\centering
		\caption{Estimated group effects from a one-way ANOVA analysis for the phthalate MEHP.} 
		\begin{tabular}{rrrrr}
			\hline
			& Estimate & Std. Error & t value & Pr($>|t|$) \\ 
			\hline
			(Intercept) & 1.58 & 0.04 & 35.44 & 0.00 \\ 
			N-H Black & 0.03 & 0.06 & 0.44 & 0.66 \\ 
			N-H White & -0.25 & 0.05 & -4.57 & 0.00 \\ 
			OH & 0.01 & 0.08 & 0.10 & 0.92 \\ 
			Other/Multi & 0.00 & 0.09 & 0.02 & 0.99 \\ 
			\hline
		\end{tabular}
		\label{last}
	\end{table}
	
	\subsection{Additional figures}
\begin{figure}
		\centering
		\subfigure[Trace plot for $Q$]{\label{fig:a }\includegraphics[width=60mm]{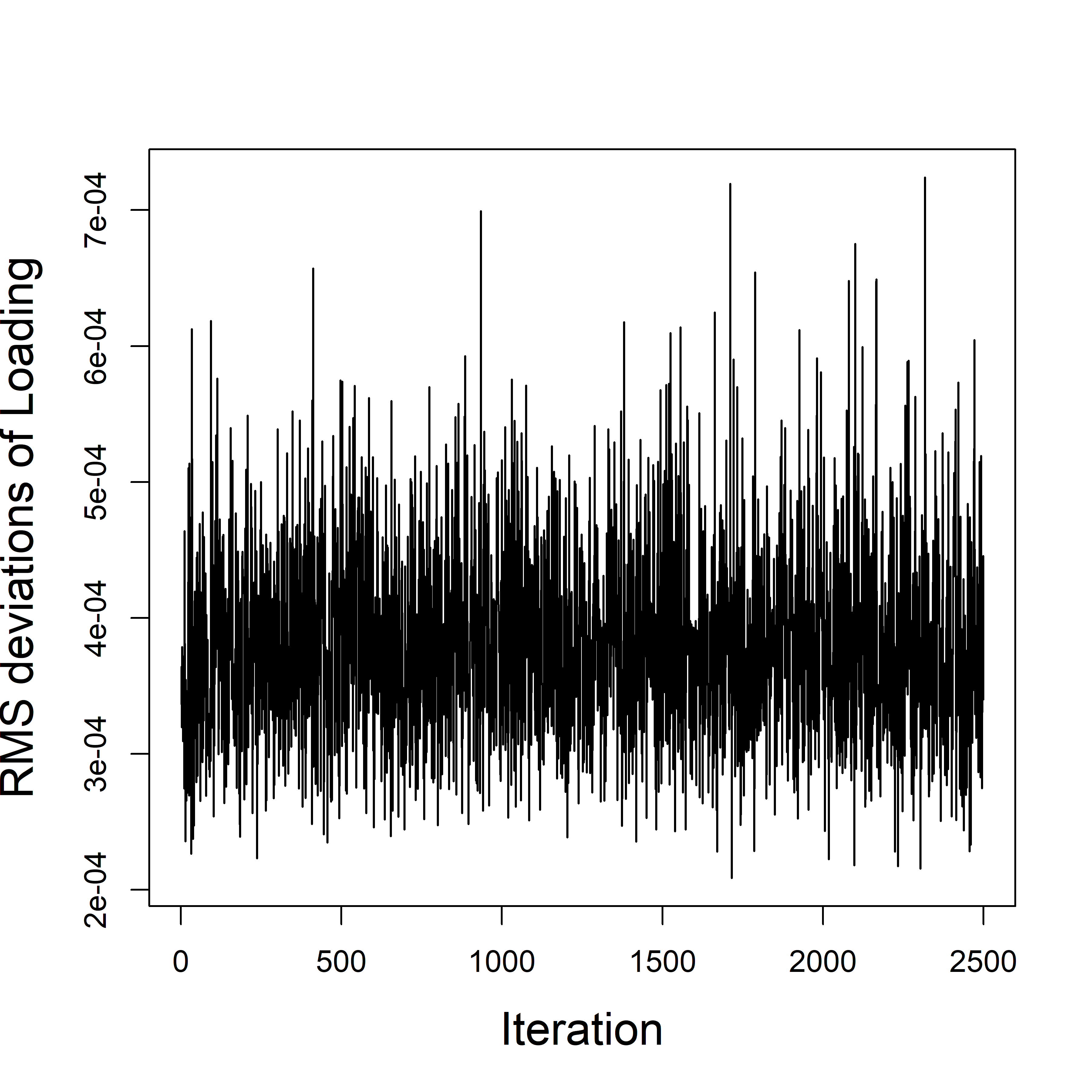}}
		\subfigure[Trace plot for $\Lambda$]{\label{fig:b }\includegraphics[width=60mm]{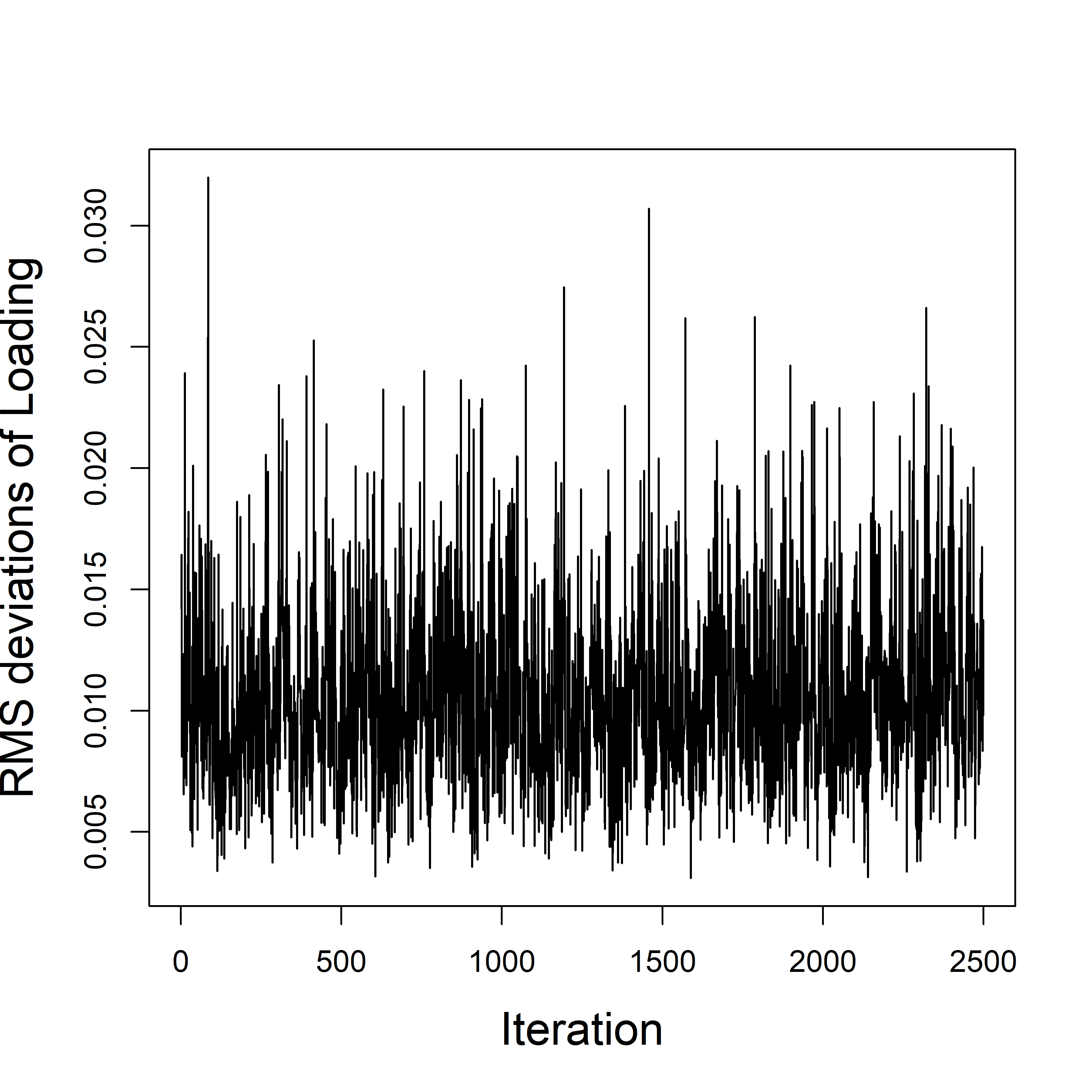}}
		\caption{Trace plots of root mean square (RMS) deviations across the MCMC chain for the perturbation matrices $\frac{1}{4}\sum_{j=2}^{5}\overline{(Q_{j}^t-Q_{j}^{t+1})^2}$ and the shared loading matrix $\Lambda$ which is $\overline{(\Lambda^t-\Lambda^{t+1})^2}$. The matrices $Q_{j}^t$ and $\Lambda^t$ are the $t$-$th$ post burn samples of $Q_{j}$ and $\Lambda$ respectively for NHANES application.}
		\label{fig:tralpha}
	\end{figure}

\begin{figure}[htbp]
		\centering
		\includegraphics[width = 1\textwidth]{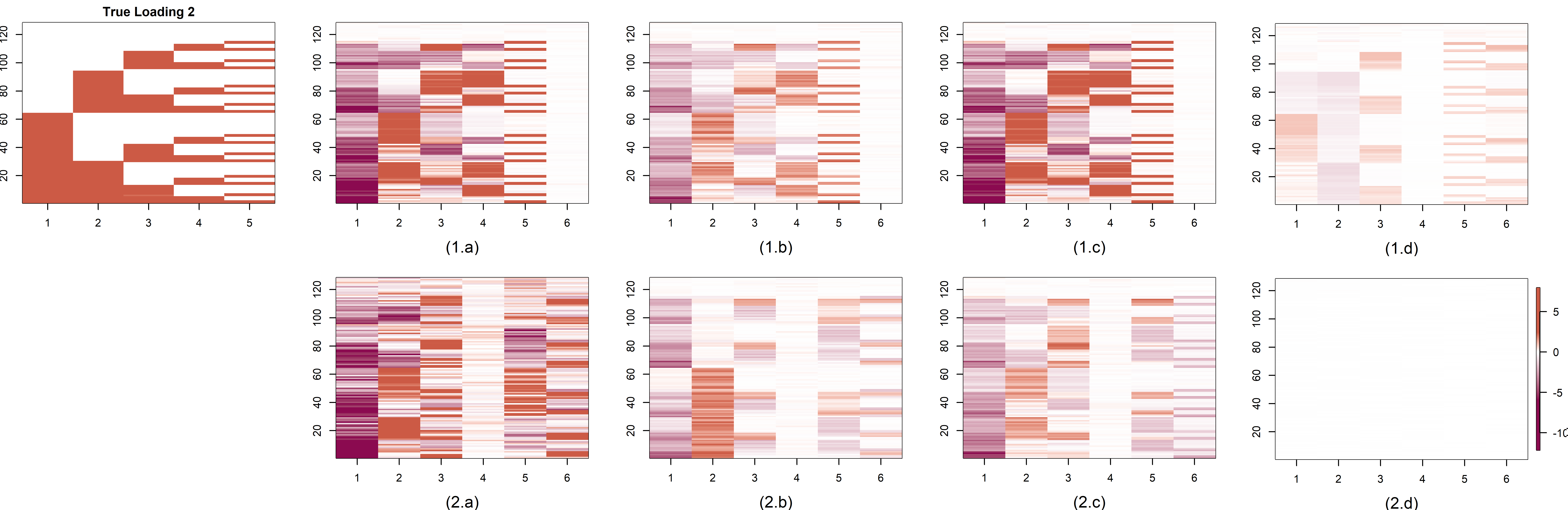}
		\caption{Comparison of estimated loading for loading matrix 2 in simulation case 2 of the manuscript with different choices of $\alpha_0$ and $\alpha$ where $Q_{j0}\sim$MN$(I_p, \alpha_0 I_p, \alpha_0 I_p)$ and $U=\alpha I_p=V.$ (a) $\alpha = 1\times 10^{-4}, \alpha_0=1\times 10^{-4}$, (b) $\alpha = 1\times 10^{-2}, \alpha_0=1\times 10^{-4}$, (c) FBPFA with $\alpha_0=1\times 10^{-4}$, (d) BMSFA with $\alpha_0=1\times 10^{-4}$, (e) $\alpha = 1\times 10^{-4}, \alpha_0=1\times 10^{-2}$, (f) $\alpha = 1\times 10^{-2}, \alpha_0=1\times 10^{-2}$, (g) FBPFA with $\alpha_0=1\times 10^{-2}$, (h) BMSFA with $\alpha_0=1\times 10^{-2}$. True loading matrices are plotted twice in columns 1 for easier comparison with other images. The color gradient added in the last image holds for all the images.}
		\label{pertothergrpother}
	\end{figure}

\begin{figure}
		\centering
		\includegraphics[width = 1\textwidth]{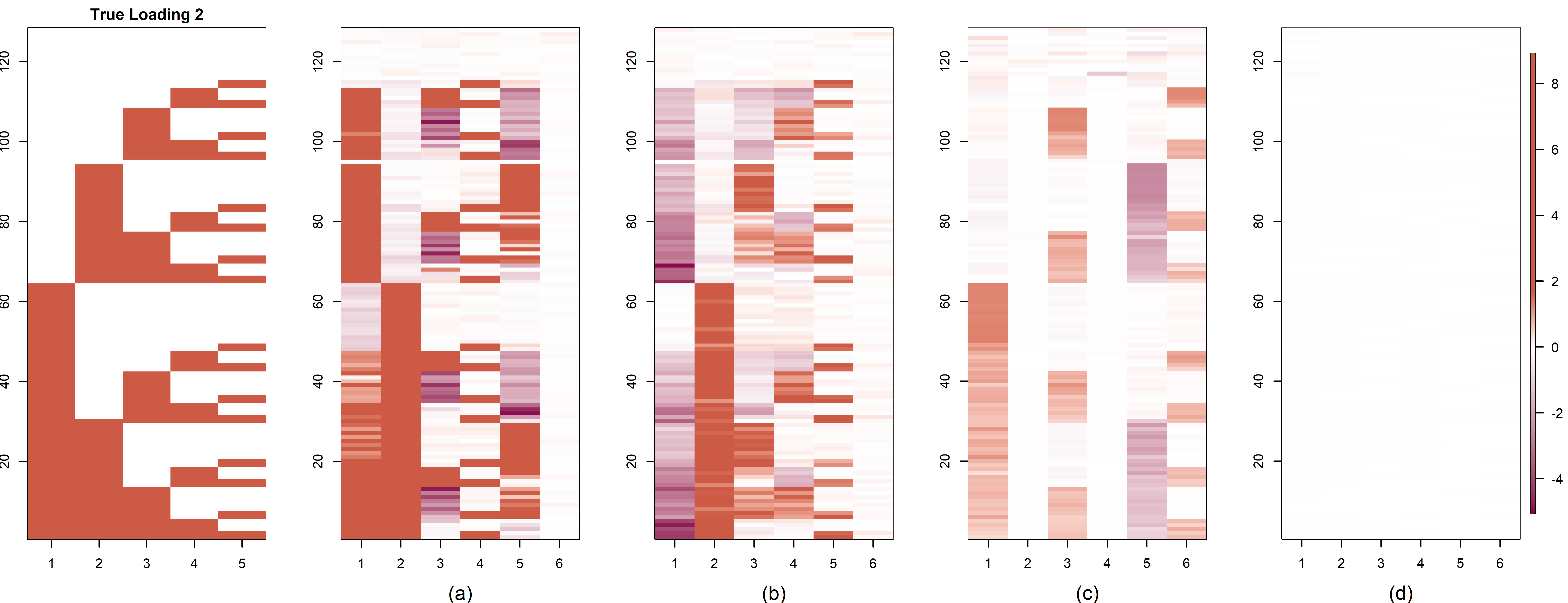}
		\caption{Comparison of estimated loading in the partially shared modification of simulation case 2 of the manuscript.  Row 1 corresponds to true loading structure 1 and row 2 to true loading structure 2. (a) FBPFA with $\alpha_0=1\times 10^{-4}$, (b) FBPFA with $\alpha_0=1\times 10^{-2}$, (c) BMSFA with $\alpha_0=1\times 10^{-4}$, (d) BMSFA with $\alpha_0=1\times 10^{-2}$, (e) FBPFA with $\alpha_0=1\times 10^{-4}$, (f) FBPFA with $\alpha_0=1\times 10^{-2}$, (g) BMSFA with $\alpha_0=1\times 10^{-4}$, (h) BMSFA with $\alpha_0=1\times 10^{-2}$. The color gradient added in the last image holds for all the images.}
		\label{fig:pertsomeother}
	\end{figure}

\begin{figure}[htbp]
		\centering
		\includegraphics[width = 1\textwidth]{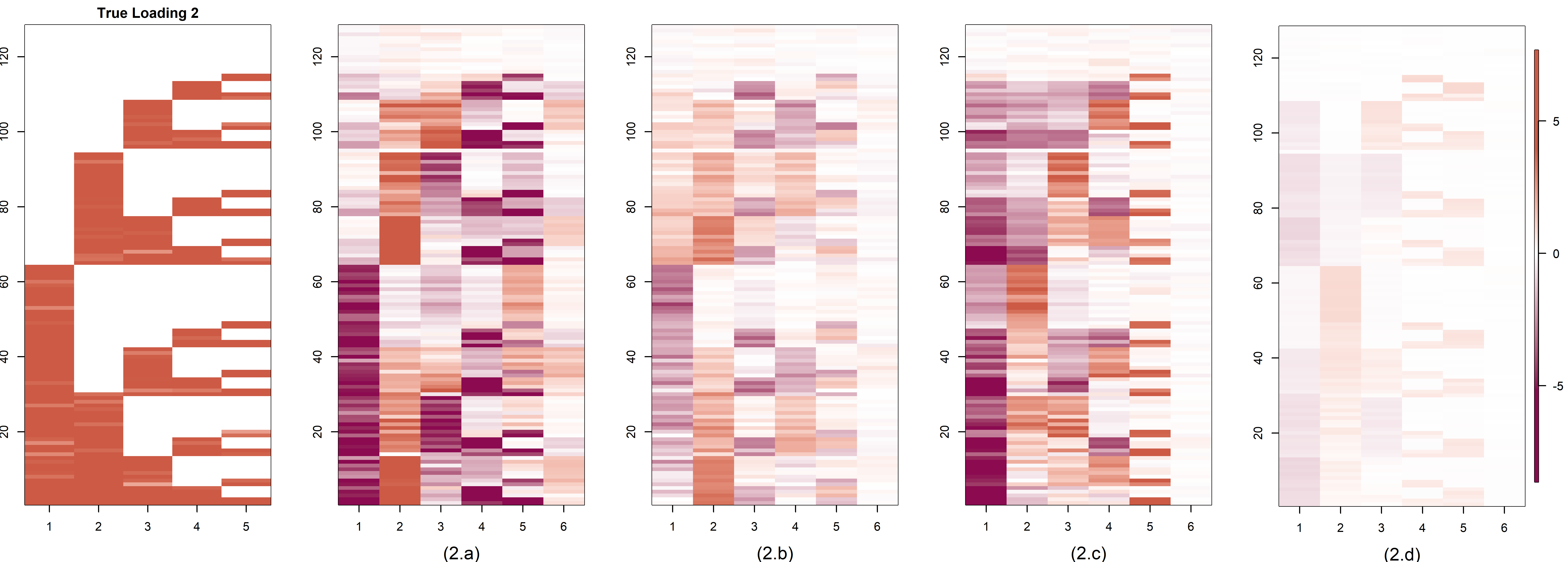}
		\caption{Comparison of estimated loading matrices in simulation case 3  of the manuscript when the true data generating process follows the model in~\eqref{BMSFA}. For the first two rows, true $\Psi_j$'s are generated from N$(-0.2, 0.2)$ and for the last two rows $\Psi_j$'s are generated from N$(-0.5, 0.8)$ and for each row (a) PFA with $\alpha = 1\times 10^{-2}$, (b)PFA with $\alpha = 1\times 10^{-4}$, (c) FBPFA, (d) BMSFA. The color gradient added in the last image holds for all the images.}
		\label{Pertnewrevother}
	\end{figure}

	We have some additional figures in this section as discussed in the Manuscript. Figure~\ref{fig:tralpha} shows the MCMC mixing of the perturbation matrices and loading matrices using FBPFA for NHANES application. Figure~\ref{pertothergrpother} illustrates estimated loadings for the second choice of loading matrix in Section 5.2 of the manuscript. From the same section, estimated loadings for the partially shared factors case are in Figure~\ref{fig:pertsomeother}. Figure~\ref{Pertnewrevother} depicts estimated loading for the same choice of loading in Section 5.3 of the manuscript.
	
	\bibliographystyle{bibstyle}
	\bibliography{main}

\end{document}